\documentclass[12pt,a4paper]{article}

\usepackage[utf8]{inputenc}
\usepackage[T1]{fontenc}
\usepackage{lmodern}
\usepackage[a4paper,left=2.85cm,right=2.85cm,top=2.55cm,bottom=2.80cm]{geometry}
\usepackage{amsmath,amssymb,mathtools}
\usepackage{booktabs,array,longtable,tabularx}
\usepackage{graphicx}
\usepackage{microtype}
\usepackage[colorlinks=true,linkcolor=blue,citecolor=blue,urlcolor=blue]{hyperref}
\usepackage[labelfont=normalfont,textfont=normalfont,font=small,labelsep=period]{caption}
\usepackage{authblk}
\usepackage{placeins}
\usepackage{float}
\usepackage{subfig}

\usepackage{xcolor}
\definecolor{darkgreen}{rgb}{0.0, 0.5, 0.0}

\graphicspath{{figures/}}
\numberwithin{equation}{section}
\allowdisplaybreaks
\newcommand{\dd}{\mathrm d}
\newcommand{\ab}{\bar\alpha_s}
\newcommand{\aw}{a_\omega}
\newcommand{\GG}{\mathcal G}
\newcommand{\cO}{\mathcal O}
\DeclareMathOperator{\diag}{diag}

\hypersetup{
  pdftitle={BFKL as a matrix: anomalous dimensions  and running-coupling},
  pdfauthor={Josep Rubi Bort, Agustin Sabio Vera, Eduardo Serna Campillo},
  pdfsubject={A finite-matrix formulation of BFKL scale evolution and spectral limits},
  pdfkeywords={BFKL evolution, virtuality lattice, anomalous dimension, running coupling, Regge cut, discrete Pomeron}
}

\title{BFKL as a matrix:\\ anomalous dimensions and running-coupling}
\author[1]{Josep Rub\'i Bort}
\author[1,2]{Agust\'in Sabio Vera\thanks{Corresponding author: \href{mailto:agustin.sabio@uam.es}{agustin.sabio@uam.es}}}
\author[1]{Eduardo Serna Campillo}
\affil[1]{Instituto de F\'isica Te\'orica UAM/CSIC, Nicol\'as Cabrera 13--15, E-28049 Madrid, Spain}
\affil[2]{Theoretical Physics Department, Universidad Aut\'onoma de Madrid, E-28049 Madrid, Spain}
\date{}

\begin{document}
\maketitle
\vspace{-1.6em}
\begin{center}
\small\itshape Dedicated to the memory of C\'esar G\'omez L\'opez
\end{center}
\vspace{0.25em}

\begin{abstract}
The BFKL equation resums high-energy logarithms in a scattering process.  
We average over the angle between transverse momenta and replace the 
squared transverse momentum (\emph{virtuality}) by $D$ discrete cells.  The evolution equation becomes a real symmetric $D \times D$
matrix $H_D$.  Its off-diagonal entries describe
real gluon emissions that move the scale from one cell to another. Its 
diagonal is the virtual correction that makes those emissions finite.  Powers of the
matrix build up ordered emission histories, the inverse matrix propagates the
amplitude between scales, and the eigenvalues and eigenvectors give
the positions and the strengths of the singularities that control growth with
rapidity.

A row far from both ends of the matrix reproduces the BFKL
eigenvalue.  Some regions bring corrections to this for a finite Hamiltonian.
These are the missing regions below the first cell and above the last cell, and the cell width.  Matching the two
scale powers allowed by a row yields the anomalous dimension, an 
exponent controlling  how the gluon distribution changes with the hard scale.
The matrix reproduces the leading-logarithmic small-$x$
coefficients.

At fixed coupling, every finite matrix has only isolated poles.  We show that
these poles become increasingly dense as the logarithmic virtuality range
grows, and we state the additional condition on their residues that is needed
for the sum to approach the continuous BFKL Regge cut.  Multiplying each real
emission by a Bessel factor, which sums the leading double logarithms of the
scale ratio, provides a controlled test of how the fixed-coupling spectrum and
its leading eigenvectors respond when long jumps in scale are suppressed.

Finally, we place a running coupling which avoids the Landau pole.  The number of positive low-scale states grows
linearly with the available logarithmic length, with a slope fixed by the
kernel.  As a separate benchmark, a standard one-loop coupling with Landau pole, combined with an imposed
lower wall, creates an ultraviolet turning point and an Airy decay region. 
\end{abstract}


\section{Why a finite-virtuality Hamiltonian}
\label{sec:intro}

When a hard probe strikes a proton at high energy, the scattering
amplitude is built from a chain of emitted gluons, and each emission can move
the transverse momentum carried along that chain either upward or downward.
Two questions are central: how fast does the amplitude grow as the
collision energy increases, and how does it change when the resolving scale of
the probe is varied?  This paper replaces the continuous range of transverse
momenta by a finite list of cells, so that both questions can be answered with
one matrix, through its powers, inverse and eigenvalues.  Throughout
this work, \emph{virtuality} means the squared transverse momentum $q^2$.

The Balitsky--Fadin--Kuraev--Lipatov (BFKL) equation describes the Regge limit,
the regime in which the collision energy is much larger than every transverse
scale that resolves the process.  The size of that separation is measured by a
rapidity interval $Y$.  For scattering at fixed momentum transfer and for
deep-inelastic scattering (DIS) it is, respectively,
\begin{equation}
Y\simeq\log\frac{s}{|t|},
\qquad
Y\simeq\log\frac1x.
\label{eq:rapidity}
\end{equation}
Here $s$ is the squared centre-of-mass energy, $t$ is the momentum transfer,
and Bjorken $x$ is the fraction of the proton momentum picked out by the hard
probe.  Rapidity measures the phase space available along the collision axis for
further radiation. Larger $Y$ gives the gluon chain more opportunities to
emit.

At order $\alpha_s^n$ the largest  term is 
$(\bar\alpha_sY)^n$, with $\bar\alpha_s=\alpha_sN_c/\pi$ and $N_c=3$ the number
of colours in QCD.  Leading-logarithmic (LL) accuracy means keeping this
largest power of $Y$ at each order in the coupling.  When $\bar\alpha_sY$ is of
order one, no fixed number of orders is enough, and the BFKL equation sums the
whole LL series
\cite{FadinKuraevLipatov:1975,KuraevLipatovFadin:1976,
KuraevLipatovFadin:1977,BalitskyLipatov:1978}.

After averaging over the relative azimuthal angle, the only variable left is
$q^2$.  The BFKL kernel  determines how the amplitude at one value of $q^2$ receives
contributions from all other values.  At fixed coupling this kernel has no
preferred transverse scale and multiplying every virtuality by the same positive
number leaves it invariant.
Throughout the rest of the paper we work in the logarithmic coordinate
\[
t \equiv \log\!\left(\frac{q^2}{\mu^2}\right),
\]
in which a rescaling of $q^2$ is an ordinary translation of $t$. The
analogue of a plane wave is therefore a power law,
\[
f_\gamma(q^2)=\left(\frac{q^2}{\mu^2}\right)^{\gamma-1},
\]
where the arbitrary scale $\mu^2$ serves only to make the ratio dimensionless.
The function $f_\gamma$ is \emph{unintegrated};  it is a density in $q^2$. Integrating it up to a hard scale $Q^2$ gives a growing part proportional to $(Q^2/\mu^2)^\gamma$.  The exponent $\gamma$ is
the anomalous dimension, that is, the logarithmic rate at which the
distribution changes with the hard scale,
\begin{equation}
\frac{\partial\log g_\omega(Q^2)}
     {\partial\log(Q^2/\mu^2)}
=\gamma(\omega).
\label{eq:anomalous_plain}
\end{equation}
The variable $\omega$ is conjugate to rapidity under a Mellin transform. It converts a function of
$Y$ into a function of $\omega$ by integrating against $e^{-\omega Y}$, and $g_\omega$ is the Mellin moment of the integrated gluon distribution.  Under
this transform, a term $e^{\lambda Y}$ becomes $1/(\omega-\lambda)$, so the
question ``how fast does the amplitude grow?'' turns into the question ``where
are the singularities in $\omega$?''

The fixed-coupling continuum spectrum is most naturally described on the
symmetric line $\gamma=\tfrac12+i\nu$.  There the two powers related by
$\gamma\leftrightarrow1-\gamma$ have the same magnitude, and $\nu$ is an
ordinary wave frequency in the variable $t$.  The largest continuum eigenvalue, reached at $\nu= 0$,
is
\begin{equation}
\omega_P=4\bar\alpha_s\log2.
\label{eq:bfkl_edge_intro}
\end{equation}
In Regge theory, the Pomeron is the leading exchange with vacuum quantum
numbers, and $\omega_P$ is the amount by which its LL fixed-coupling intercept exceeds one, $\alpha_P(0)-1$.  The difference between a pole and a cut can be stated directly in
rapidity space.  A single growth rate $\lambda$ gives one exponential and one
pole,
\[
e^{\lambda Y}
\quad\longleftrightarrow\quad
\hbox{a pole at }\omega=\lambda,
\]
whereas a continuum of rates reaching up to $\omega_P$ gives
\[
\frac{e^{\omega_PY}}{\sqrt{Y}}
\quad\longleftrightarrow\quad
\hbox{a branch cut ending at }\omega_P.
\]
The extra factor $Y^{-1/2}$ accounts for the shrinking window of nearby modes that
still contribute at large $Y$.  Section~\ref{sec:fixed_spectrum} derives this
factor and shows how a finite list of poles approaches the BFKL Regge cut
\cite{ForshawRoss:1997,Collins:1977}.
The Green function of the BFKL equation admits the spectral decomposition
\begin{equation}
	G(\omega, q, q^\prime) = \int \frac{\phi(\nu, q) \phi^*(\nu, q^\prime)}{\omega - \epsilon(\nu)}\dd\nu,
\end{equation}
where $\epsilon(\nu)$ are the eigenvalues of the kernel. The most studied regime of the BFKL equation is that in which $q\sim q^\prime$. There are other useful regimes, such as the one where $q \gg q^\prime$. There, the Dokshitzer--Gribov--Lipatov--Altarelli--Parisi (DGLAP) evolution governs the parton distribution functions. In this limit, BFKL results such as the small-$x$ anomalous dimension can be compared with DGLAP \cite{ballAllOrderRunning2006}.

The aim of this paper is to use a discretised version of the BFKL equation
to reproduce continuum results and to identify  the
limits in which they are recovered. Our starting point is the framework of
Refs.~\cite{Bethencourt:2020,chachamis2023vonneumannentropylindblad}, where
the discretisation of the BFKL equation and its continuum limit were
introduced and connected to the density operator, the von Neumann entropy
and decoherence. Here we extend that framework in several directions, each
section building on the previous one. Section~\ref{sec:finite} derives the
finite Hamiltonian and its Green function. Section~\ref{sec:anomalous}
obtains the continuum characteristic function from a single row, through an
exact boundary-retaining identity with a controlled finite-row error, and
connects it to the source-to-observation exponent.
Section~\ref{sec:numerics} tests the three-scale limit
$1\ll N_0\ll N\ll D$, from which the small-$x$ DGLAP anomalous dimension is
recovered order by order. Section~\ref{sec:fixed_spectrum} finds numerical evidence for the
$1/(\log D)^2$ convergence of the spectral edge, shows how the BFKL cut
emerges from a sum of weighted poles, and studies a Bessel deformation of
the kernel motivated by collinear contributions. Finally,
Section~\ref{sec:running} introduces one-loop running through the coupling
and compares an infrared-completed matrix with an infrared wall, using the
turning point as a benchmark.
\section{The matrix and its Green function}
\label{sec:finite}

Now we derive the entries of the Hamiltonian and then use the
matrix to propagate an amplitude between two cells.  Once the cell convention
has been fixed, every identity in this section is exact at finite $D$ for $\mathcal{H}_D$; $H_D$ is its fixed$-i$, $D\to \infty$ limit.

\subsection{Real emission and virtual subtraction}

The azimuthal angle is the direction of a transverse momentum around the beam
axis.  Averaging over it keeps only the angle-independent component and
discards the information about that direction. The azimuthally averaged ($n=0$) BFKL equation can be written as
\begin{equation}
\begin{aligned}
\frac{\partial f(q_1^2,Y)}{\ab\,\partial Y}
={}&
\int_0^\infty
\frac{\dd q_2^2}{|q_2^2-q_1^2|}
\bigl[f(q_2^2,Y)-f(q_1^2,Y)\bigr]
+
\int_{q_1^2}^\infty
\frac{\dd q_2^2}{q_2^2}f(q_1^2,Y).
\end{aligned}
\label{eq:continuum_kernel}
\end{equation}
Here $q_1^2$ is the virtuality at which the evolved amplitude is observed,
while $q_2^2$ is integrated over in one application of the kernel.

We now divide the positive virtuality axis into cells of equal width,
\[
q_2^2=j\Delta,
\qquad j=1,2,\ldots, \qquad dq_2^2 = \Delta, \qquad q_1^2 = i\Delta.
\]

Row $i$ describes the evolution at the output virtuality $q_1^2$, and column
$j$ labels the virtuality from which a contribution arrives.  For two different cells,
\begin{equation}
\Delta\frac{1}{|i\Delta-j\Delta|}
=
\frac{1}{|i-j|}.
\label{eq:cell_hop}
\end{equation}
Small changes of scale are favoured.

The diagonal entry follows from discretising the two virtual subtractions.  For
a fixed row $i$, the cells below it give
\begin{equation}
-\sum_{j=1}^{i-1}\frac{1}{i-j}
=
-\sum_{r=1}^{i-1}\frac1r
=
-h_{i-1},
\qquad
h_m=\sum_{r=1}^{m}\frac1r,
\qquad
h_0=0,
\label{eq:lower_virtual}
\end{equation}
where $h_m$ is the $m$-th harmonic number. 

\begin{figure}[ht]
	\centering
	\includegraphics[width=0.7\textwidth]{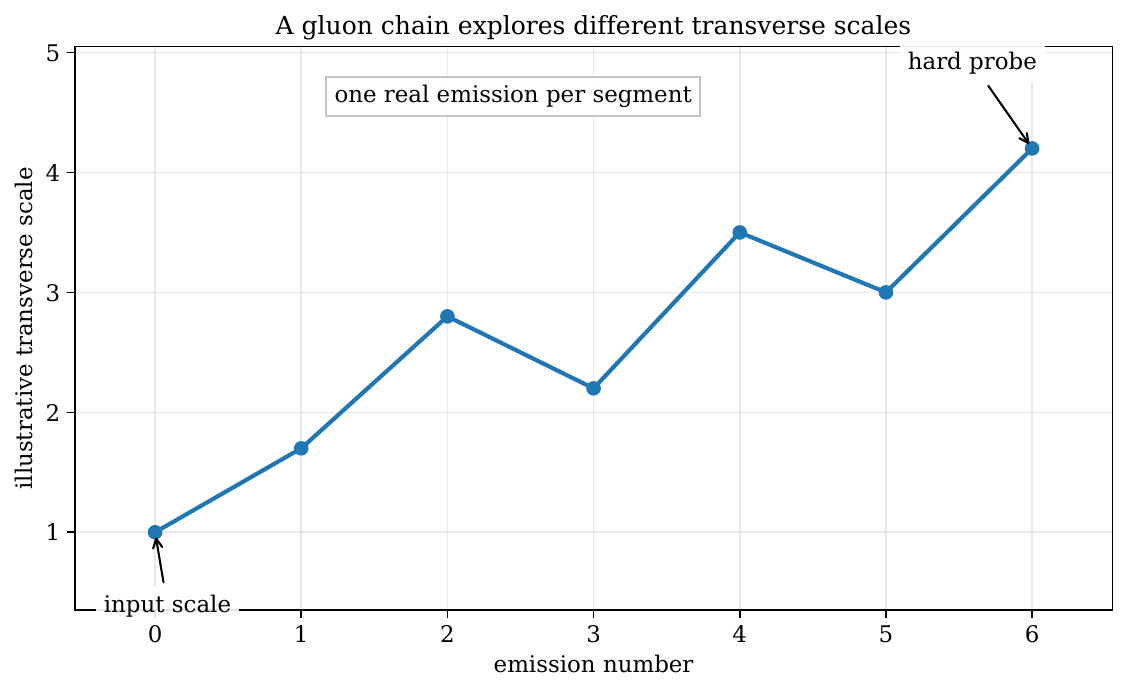}
	\caption{Schematic high-energy emission chain.  The hard probe resolves the
		process at a scale $Q^2$, while the target supplies a lower input scale
		$Q_0^2$.  Successive emissions may raise or lower the transverse virtuality.}
	\label{fig:physical_chain}
\end{figure}

For the cells above $i$, we retain the upper cells $i\leq j \leq D$ and let $D\to\infty$ at the end.
Including the cell that touches the lower limit of the last integral, with the
same discretisation convention, the upper contribution is
\begin{equation}
\begin{aligned}
\frac1i+\sum_{j=i+1}^{D}
\left(-\frac{1}{j-i}+\frac1j\right)
=
\frac1i-h_{D-i} + h_{D} - h_{i}
=
-h_{i-1}+h_{D}-h_{D-i}.
\end{aligned}
\label{eq:upper_virtual}
\end{equation}
The first term $1/i$ comes from that boundary cell.  This convention is fixed
once and for all.\footnote{Starting the sum one cell higher would shift the
diagonal to $-h_{i-1}-h_i=-2h_{i-1}-1/i$.  The shift lies inside the regular
$\cO(i^{-1})$ row error, but it changes every exact finite identity.}
For fixed $i$,
\begin{equation}
	\label{Eq: Harmonic difference approximation}
	h_{D}-h_{D-i}
	=
	\sum_{r=0}^{i-1}\frac{1}{D-r}
	\longrightarrow0
	\qquad (D\to\infty),
\end{equation}
so the upper cells provide a second contribution $-h_{i-1}$.  Adding the lower
and upper parts gives the diagonal entry $-2h_{i-1}$, Eq.~\eqref{eq:main_matrix} below.  The two copies of
the harmonic number have different origins: one comes from virtualities below
the output scale, the other from the real--virtual cancellation above it. Eq.~\eqref{Eq: Harmonic difference approximation} motivates us to write the kernel in two different ways. 
The first, more common in the literature~\cite{Bethencourt:2020,chachamis2023vonneumannentropylindblad}, reads
\begin{equation}
	(H_D)_{ij}
	=
	\frac{1-\delta_{ij}}{|i-j|}
	-2h_{i-1}\delta_{ij}.
	\label{eq:main_matrix}
\end{equation}
However, in Sect.~\ref{sec:anomalous} we will see that keeping Eq.~\eqref{Eq: Harmonic difference approximation} will prove useful,
\begin{equation}
	\label{Eq: H exact}
	(\mathcal{H}_D)_{ij} = \frac{1-\delta_{ij}}{|i-j|} + \left(- 2h_{i-1} + h_D - h_{D-i} \right)\delta_{ij}.
\end{equation}
We use this form in Sect.~\ref{subsect: Char funct} and Eq.~\eqref{eq:main_matrix} in the rest of the paper.
The two differ only by the boundary term $h_D-h_{D-i}$, which vanishes for fixed $i$ as $D\to\infty$. A positive off-diagonal entry is a real emission that changes the
scale. The negative harmonic diagonal is the virtual subtraction that cancels the local singularity of real emission.  Both matrices are real and symmetric, so their eigenvalues are real and their eigenvectors can be chosen orthonormal.
Fig.~\ref{fig:physical_chain} gives the physical picture behind the
construction.  A sequence of emissions changes the transverse scale carried by
the exchanged state, so the evolution must keep track of both the energy
interval and the path through virtuality.
We call $H_D$ and $\mathcal{H}_D$ Hamiltonians because they generate evolution in rapidity in the
same algebraic way that a Hamiltonian generates evolution in time.

The power $H_D^r$ adds up all ordered histories containing r interactions. The inverse $(1 - a_\omega H_D)^{-1}$, with $a_\omega = \bar{\alpha}_s/\omega$, is the Green function: place a unit source in one cell and it returns the amplitude that arrives in another cell after any number of interactions.  The eigenvalues give the positions of
the poles, and the eigenvectors give their residues, that is, the
strengths with which each mode couples to the chosen source and observation
cells.  A running coupling will later be included by dressing the same matrix
with scale-dependent weights.

To obtain our results we will require three limits, two of which are necessary for the continuum limit, and the third necessary to connect with the DGLAP anomalous dimension. They are given by
\begin{equation}
	1\ll N_0\ll N\ll D.
	\label{eq:three_scales}
\end{equation}
Here $N_0$ is the source cell, $N$ the observation cell, and $D$ the last
(ultraviolet) cell. \begin{figure}[ht]
	\centering
	\includegraphics[width=0.75\textwidth]{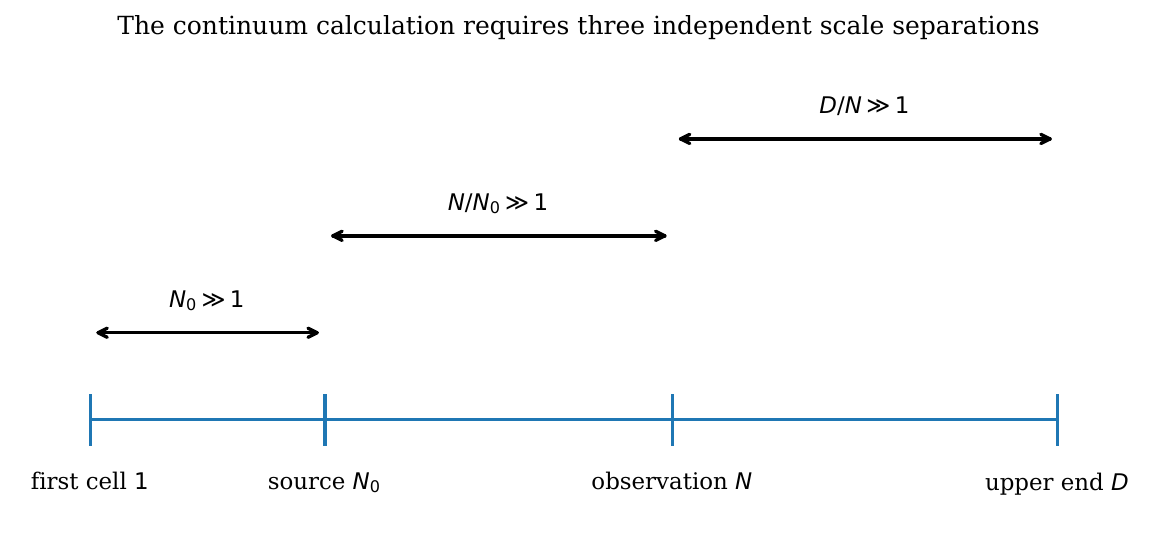}
	\caption{The three independent scale separations of
		Eq.~\eqref{eq:three_scales}.  Moving $N_0$ removes the lower-boundary effect,
		increasing $N/N_0$ exposes the long scale logarithm, and increasing $D/N$
		restores the real-emission region above the observation cell.}
	\label{fig:three_scale_geometry}
\end{figure}
The second separation is not required to obtain the continuum limit, but to obtain the expansion of the anomalous dimension, as discussed in the introduction. BFKL studies usually consider comparable scale (i.e., $q\sim q^\prime$, equivalently $N_0\sim N$). This second separation, $N_0 \ll N$, creates a long interval $\log(N/N_0)$ over which a scale
exponent can be measured.  The final condition keeps the real-emission tail above the
observation scale.  No increase in matrix size or in arithmetic precision can
substitute for a missing one of these three geometrical separations. The basic structure behind these limits can be seen in Fig.~\ref{fig:three_scale_geometry}. Each inequality in Eq.~\eqref{eq:three_scales} removes a
different finite-lattice effect.

The upper boundary condition used so far is known as the \emph{finite-box} prescription. However, there are other methods that give the Hamiltonian a different interpretation. Although it is not the focus of this work, one alternative deserves mention. The most interesting one is the \emph{principal-section} prescription, which places the upper boundary at the row-dependent cell $R+i$ instead of the fixed cell $D$. Both prescriptions are connected through $D = R+i$. The diagonal part of the new prescription is given by
\begin{equation}
	(\mathcal{H}_{R}^{\text{PS}})_{ij} = \frac{1 - \delta_{ij}}{|i-j|} + (h_{R+i} - h_{R} - 2h_{i-1})\delta_{ij}. 
\end{equation}
Note that if the vanishing boundary terms are dropped (Eq.~\eqref{Eq: Harmonic difference approximation}), both prescriptions yield the same diagonal, $-2h_{i-1}$.
\subsection{Evolution, inverse matrix and  poles}

A Green function answers a source-response question: if a unit amplitude starts
in one cell, how much of it reaches another cell after every allowed
interaction?  Let $|\phi(Y)\rangle$ collect the amplitudes in the $D$ cells.
Its rapidity evolution is
\begin{equation}
\frac{\partial}{\ab\,\partial Y}|\phi(Y)\rangle=H_D|\phi(Y)\rangle.
\label{eq:rapidity_matrix}
\end{equation}
A Laplace--Mellin transform weights this evolution by $e^{-\omega Y}$.  It
turns the derivative with respect to $Y$ into multiplication by $\omega$, so
the differential equation becomes
\begin{equation}
|\widetilde\phi(\omega)\rangle
=
\frac1\omega\left(1-\frac{\ab}{\omega}H_D\right)^{-1}|\phi(0)\rangle.
\label{eq:mellin_solution}
\end{equation}
Throughout we write
\begin{equation}
\aw=\frac{\ab}{\omega}.
\label{eq:aomega}
\end{equation}
We place the initial state at cell $N_0$ and take $N>N_0$, as appropriate for
evolution from an input scale to a harder observation scale.  The first
non-zero term of the inverse matrix is then
\[
\aw\langle N|H_D|N_0\rangle
=
\frac{\aw}{N-N_0}.
\]
Normalizing by this leading term defines
\begin{equation}
\GG_{D,N,N_0}(\aw)
=
\frac{N-N_0}{\aw}
\left\langle N\left|\frac{1}{1-\aw H_D}\right|N_0\right\rangle
=1+\cO(\aw),
\label{eq:normalised_green}
\end{equation}
our central source-to-observation quantity.  Expanding the inverse
matrix gives
\begin{equation}
\GG=1+\sum_{r\ge1}\aw^r M_r,
\qquad
M_r=(N-N_0)\langle N|H_D^{r+1}|N_0\rangle.
\label{eq:raw_coefficients}
\end{equation}
Each product of matrix entries sums an ordered history through intermediate
virtualities. These coefficients contain repeated products of
lower-order evolution, which are removed by expanding the logarithm,
\begin{equation}
\log\GG=\sum_{r\ge1}\aw^r C_r,
\qquad
C_r=M_r-\frac1r\sum_{s=1}^{r-1}sC_sM_{r-s},
\quad M_0=1.
\label{eq:connected_recursion}
\end{equation}
For example, $C_1=M_1$ and $C_2=M_2-M_1^2/2$.

The same inverse matrix has a history representation, valid inside its
convergence domain and order by order around $\aw=0$.  Split the
real and virtual parts,
\begin{equation}
H_D=R_D-2V_D,
\qquad
(R_D)_{ij}=\frac{1-\delta_{ij}}{|i-j|},
\qquad
(V_D)_{ij}=h_{i-1}\delta_{ij}.
\label{eq:real_virtual_split}
\end{equation}
A path $\mathcal P=(m_0,\ldots,m_J)$ starts at $m_0=N_0$, ends at $m_J=n$, and
changes cell at every step.  Summing any number of virtual insertions at each
visited cell gives
\begin{equation}
\left[(1-\aw H_D)^{-1}\right]_{nN_0}
=
\sum_{J\ge0}\aw^J
\sum_{\substack{\mathcal P:N_0\to n\\m_{s+1}\ne m_s}}
\left[\prod_{s=0}^{J-1}(R_D)_{m_{s+1}m_s}\right]
\prod_{s=0}^{J}\frac{1}{1+2\aw h_{m_s-1}}.
\label{eq:path_resolvent}
\end{equation}
The first product accounts for the real changes of cell; the second sums all local
virtual insertions at every visit.  The path formula and the eigenmode formula
below are two ways of organising the same inverse matrix: one keeps the
history, the other the normal modes. This summation over virtual insertions is the discrete analogue of gluon reggeization.
\begin{figure}[ht]
	\centering
	\includegraphics[width=0.75\textwidth]{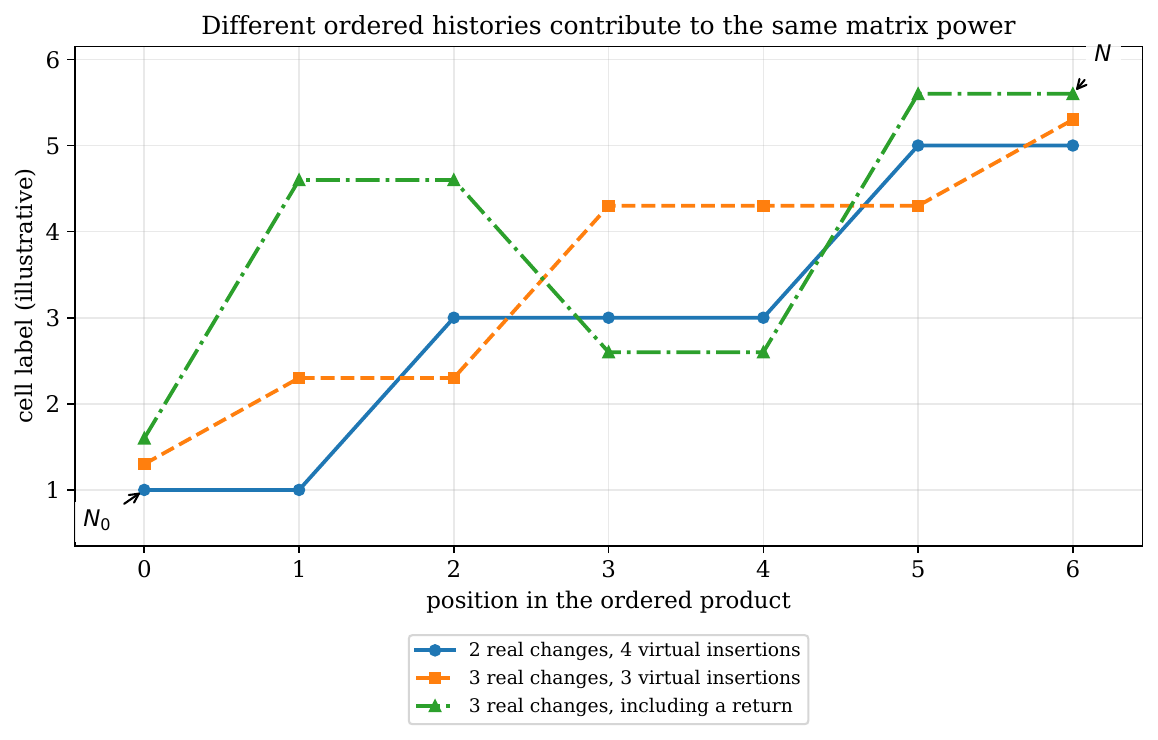}
	\caption{Examples of histories contained in matrix powers.  The paths 
		show how the products of matrix entries in
		Eq.~\eqref{eq:path_resolvent} distinguish different orderings of real and
		virtual interactions.  }
	\label{fig:path_histories}
\end{figure}  
Fig.~\ref{fig:path_histories} illustrates this formula.  A sloping
segment changes the cell and contributes a real-emission factor; a horizontal
segment leaves the cell unchanged and contributes the local virtual factor.

The same picture explains why the logarithm of the Green function is the useful
object.  Long histories can contain repeated pieces of shorter evolution, and
exactly those products appear when an exponential of a scale exponent is
expanded.  The connected coefficients $C_r$ remove them, leaving the part that
can contribute linearly to the long logarithm $\log(N/N_0)$.

The inverse matrix can also be written in the eigenbasis.  If
$H_D\phi_p=\varepsilon_p\phi_p$ with orthonormal eigenvectors, then
\begin{equation}
\left[(1-\aw H_D)^{-1}\right]_{ij}
=
\sum_{p=1}^{D}
\frac{\phi_p(i)\phi_p(j)}{1-\aw\varepsilon_p}.
\label{eq:spectral_resolvent}
\end{equation}
At finite $D$ this is a sum of simple poles, the eigenvalue fixes the
position of a pole and the product $\phi_p(i)\phi_p(j)$ is its residue for a 
pair of cells.  A Regge cut can therefore appear only once both the pole
positions \emph{and} their residues develop a continuum limit.

At fixed coupling the same representation gives
\[
[e^{\bar\alpha_sYH_D}]_{ij}
=\sum_p\phi_p(i)\phi_p(j)e^{\lambda_pY},
\qquad \lambda_p=\bar\alpha_s\varepsilon_p,
\]
so each pole is one exponential in rapidity.
Section~\ref{sec:fixed_spectrum} derives the extra $Y^{-1/2}$ factor produced
when the levels and their residues merge into a continuum.

\section{From one row to the anomalous dimension}
\label{sec:anomalous}

The next question is whether a single row of a large matrix reproduces
the scale-invariant continuum kernel.  We separate the lower-end, upper-tail
and finite-spacing errors, and then ask whether the local exponent also
controls propagation from the source cell to the observation cell.

\subsection{The continuum characteristic function}\label{subsect: Char funct}

Suppose that a row of the matrix lies far from both the first and the last
cell.  Does that row recover the BFKL kernel?  To find out, act
with the matrix on the trial power
\begin{equation}
	\label{Eq: def of w}
w_m=m^{\gamma-1},
\qquad
0<\Re\gamma<1,
\end{equation}
where the restriction on $\Re\gamma$ makes the lower- and upper-virtuality
integrals finite.  Then divide the output of row $j$ by the
value of the same trial power at that row:
\begin{equation}
\chi_{j,D}(\gamma)
=
\frac{(\mathcal{H}_Dw)_j}{w_j}.
\label{eq:row_ratio}
\end{equation}
If the finite row acted exactly as a scale-invariant operator, this ratio would
not depend on $j$ at all.  Substituting Eq.~\eqref{Eq: H exact} gives
\begin{equation}
\chi_{j,D}(\gamma)
=
\sum_{m=1}^{j-1}
\frac{(m/j)^{\gamma-1}}{j-m}
+
\sum_{m=j+1}^{D}
\frac{(m/j)^{\gamma-1}}{m-j} +
\left(h_D - h_{D-j} -2h_{j-1}\right).
\label{eq:row_raw}
\end{equation}
The first sum collects transitions towards lower virtuality and the second
those towards higher.   The BFKL kernel is finite after real emission and virtual subtraction
are combined.

To make the cancellation explicit add and
subtract the harmonic terms generated below and above row $j$.  This gives
\begin{equation}
\begin{aligned}
\chi_{j,D}(\gamma)
={}&
\sum_{m=1}^{j-1}
\frac{(m/j)^{\gamma-1}-1}{j-m}
+
\sum_{m=j+1}^{D}
\frac{(m/j)^{\gamma-1}-j/m}{m-j} + \frac{1}{j}.
\end{aligned}
\label{eq:row_subtracted}
\end{equation}
Every numerator now vanishes when $m$ approaches $j$, so the cancellation is
visible term by term, and Eq.~\eqref{eq:row_subtracted} holds for every finite
$j$ and $D$. Because Eq.~\eqref{Eq: H exact} retains the boundary term, no diagonal remainder survives. With $H_D$, the term $h_D-h_{D-j}$ would appear explicitly. This is the reason for keeping it.
Set $x=m/j$.  When both $j\to\infty$ and $D/j\to\infty$, the sums become
integrals over virtuality ratios and give
\begin{equation}
\chi(\gamma)
=
\int_0^1
\frac{x^{\gamma-1}+x^{-\gamma}-2}{1-x}\,\dd x
=
2\psi(1)-\psi(\gamma)-\psi(1-\gamma).
\label{eq:chi}
\end{equation}
Here $\psi(z)=\Gamma'(z)/\Gamma(z)$ is the digamma function, the derivative of
$\log\Gamma(z)$.  It has a simple
pole at the origin, and it grows logarithmically at large argument. The factor
$x^{\gamma-1}$ comes from motion towards lower virtuality, $x^{-\gamma}$ from
motion towards higher virtuality, and $-2$ is the continuum form of the two
virtual subtractions.  Exchanging the two directions sends $\gamma$ to
$1-\gamma$, so $\chi(\gamma)=\chi(1-\gamma)$.

The limit $\gamma\to0$ is called the collinear limit; it describes strongly
ordered evolution towards virtualities much smaller than the observation scale.
There
\begin{equation}
	\label{Eq: Collinear pole}
	\int_0^1x^{\gamma-1}\,\dd x=\frac1\gamma,
\end{equation}
while the upper-virtuality contribution stays finite.   The same region generates large
double logarithms beyond LL accuracy, which is the reason for the Bessel
deformation introduced in Section~\ref{sec:fixed_spectrum}.

To approach the continuum results, a large matrix is not enough by itself.  Row $j$ must be far from the lower
boundary, and the final cell $D$ must be far above that row.  Estimating the
two omitted boundary intervals gives
\begin{equation}
\chi_{j,D}(\gamma)-\chi(\gamma)
=
\cO\!\left(\frac{j^{-\Re\gamma}}{\Re\gamma}\right)
+
\cO\!\left[\frac{(j/D)^{1-\Re\gamma}}{1-\Re\gamma}\right]
+
\cO(j^{-1}).
\label{eq:row_error}
\end{equation}
These terms have different origins.  $j^{-\Re\gamma}/\Re\gamma$  comes from the unresolved lower interval: $0<x<1/j$, (i.e. $m = \cO(1)$)  gives
$\int_0^{1/j}x^{\Re\gamma-1}\dd x$. In other words, this is the interval below the first lattice cell that the lattice cannot resolve.
$(j/D)^{1-\Re\gamma}/(1-\Re\gamma)$ comes from the real-emission tail above the final cell: $x>D/j$ (i.e. $m > D$) gives
$\int_{D/j}^{\infty}x^{\Re\gamma-2}\dd x$.
$j^{-1}$ is the error from replacing a sum by its integral, together with the two cells adjacent to $1-1/j < x < 1$ and $1 < x < 1+1/j$ (i.e. $|m-j| = \cO(1)$), each of which contributes at $\cO(j^{-1})$.

Eq. (\ref{eq:row_error}) is a fixed-$\gamma$ estimate, with constants uniform for $\epsilon\le\Re\gamma\le1-\epsilon$ and bounded $\Im\gamma$.
The first correction matters most on the perturbative branch, where $\gamma$
starts near zero and $j^{-\gamma}$ decreases very slowly.  Moving the
ultraviolet boundary farther away cannot remove it.

The corrections in Eq.~\eqref{eq:row_error} can be separated by
changing one part of the geometry at a time.
\begin{figure}[ht]
	\centering
	\includegraphics[width=0.7\textwidth]{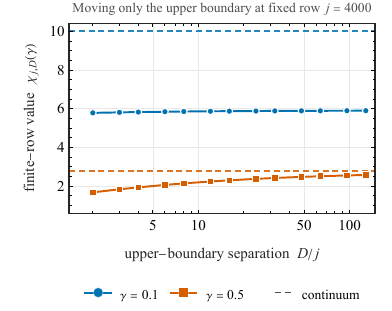}
	\caption{Upper-tail test of Eqs.~\eqref{eq:row_raw}--\eqref{eq:row_error}.
		The row is fixed at $j=4000$, $\gamma=0.1$ and $0.5$, and $D/j$ takes
		$2,3,4,6,8,12,16,24,32,48,64,96,128$.  The dotted horizontal lines are the
		continuum values from Eq.~\eqref{eq:chi}.}
	\label{fig:row_upper_tail}
\end{figure}
Fig.~\ref{fig:row_upper_tail} holds the row fixed at $j=4000$ and moves only
the upper boundary.  Increasing $D/j$ restores more of the high-virtuality tail
without improving the resolution of the interval below the first cell.  For $\gamma=1/2$ the
lower correction is already modest and the finite row moves visibly towards
$\chi(1/2)=4\log2$.  For $\gamma=0.1$ the sequence flattens while it is still
far below $\chi(0.1)$.  This is not a failure of the ultraviolet limit: the remaining error is dominated by the lower-boundary term $j^{-0.1}/0.1$, which does
not change when $D$ is enlarged at fixed $j$.

\begin{figure}[ht]
\centering
\includegraphics[width=0.7\textwidth]{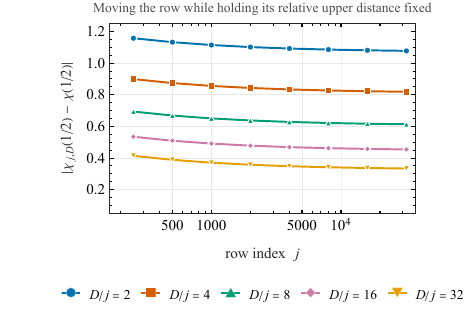}
\caption{Lower-end and spacing test of Eq.~\eqref{eq:row_error}.  Here
$\gamma=1/2$, $j=250,500,1000,2000,4000,8000,16000,32000$, and each curve
keeps one of $D/j=2,4,8,16,32$ fixed.  The vertical axis is the
absolute difference from the continuum value $\chi(1/2)=4\log2$.  Moving the
row and moving the final ultraviolet boundary test different corrections.}
\label{fig:row_lower_spacing}
\end{figure}

Fig.~\ref{fig:row_lower_spacing} performs the complementary test.  Along each
curve the ratio $D/j$ is fixed while $j$ increases.  This reduces the
lower-boundary and finite-spacing errors, but it leaves the relative
ultraviolet error unchanged because $j/D$ is constant.  Each sequence
therefore approaches a residual value set by its chosen $D/j$.  The residual
falls when that ratio is increased from $2$ to $32$.  The figure shows
that the limits $j\to\infty$ and $D/j\to\infty$ are independent.

Figs.~\ref{fig:row_upper_tail}--\ref{fig:row_lower_spacing}
show that the continuum characteristic function is recovered only through a
two-sided limit.  The row must move far above the first cell, and the
ultraviolet boundary must recede independently.  They also explain why the
perturbative region near $\gamma=0$ requires especially large scale
separations.

We can now connect this local row result to the Green function.  Let
\begin{equation}
u_j
=
\left\langle j\left|
\frac{1}{1-\aw H_D}
\right|N_0\right\rangle
\end{equation}
be the $j$-th column of the inverse matrix generated by the source cell $N_0$.  The
defining equation $(1-\aw H_D)u=|N_0\rangle$ implies, for a row $j$ away from
the source, that $u_j=\aw(H_Du)_j$.  If the solution is locally proportional to
$j^{\gamma-1}$, the continuum row limit gives (see what was done starting from Eq.~\eqref{Eq: def of w})
\begin{equation}
1=\aw\chi(\gamma).
\label{eq:characteristic}
\end{equation}
This equation fixes the local scale exponent associated with the Mellin
parameter $\aw$.  The next step is global: we must explain why the same
exponent controls propagation across the long interval from $N_0$ to $N$.

\subsection{Matching the two scale powers}

Equation~\eqref{eq:characteristic} has two solutions related by
$\gamma\leftrightarrow1-\gamma$, with scale powers
$\left(q^2/\mu^2\right)^{\gamma-1}$ and $\left(q^2/\mu^2\right)^{-\gamma}$.
The situation is the familiar one of a second-order boundary-value problem: one
independent solution is selected by regularity at the lower end, the other by
decay at the upper end, and the propagator between two interior points is built
from both.  The arbitrary powers of $\mu^2$ are absorbed into the boundary
matching factor and disappear from the final ratio $N/N_0$.  On the
perturbative branch $0<\gamma<1/2$, a Green function in a long, approximately
scale-invariant interval uses the less singular power near the lower end and
the faster decaying power near the upper end, so its middle-region form is
\begin{equation}
\left\langle N\left|\frac{1}{1-\aw H_D}\right|N_0\right\rangle
\sim
C(\aw)\,N_0^{-\gamma}N^{\gamma-1}.
\label{eq:two_power_match}
\end{equation}
The matching coefficient $C$ depends on the two boundary regions but does not grow with the
length of the middle interval.  Multiplying by the normalisation in
Eq.~\eqref{eq:normalised_green} gives
\begin{equation}
\GG_{D,N,N_0}(\aw)
\sim
\widetilde C(\aw\mid D,N_0,N)
\left(\frac{N}{N_0}\right)^{\gamma(\aw)}.
\label{eq:green_power}
\end{equation}
The explicit factor $N-N_0\simeq N$ converts the unintegrated power
$N^{\gamma-1}$ into the integrated scale power $N^\gamma$.  This is the central
bridge of the paper, because it joins two quantities that are defined in
different ways.  The exponent $\gamma$ entered through
Eq.~\eqref{eq:characteristic}, an algebraic condition on a single interior row,
which by itself says nothing about propagation between distant cells.  The
anomalous dimension on the left of Eq.~\eqref{eq:anomalous_plain} is instead a
derivative of the evolved distribution with respect to the hard scale.
Equation~\eqref{eq:green_power} identifies the two.  Since $q_N^2=N\Delta$, the
variable $\log N$ is $\log Q^2$ up to a constant, and
\[
\frac{\partial\log\GG_{D,N,N_0}}{\partial\log N}
=
\gamma(\aw)+\frac{\partial\log\widetilde C}{\partial\log N},
\]
so the local row exponent becomes the hard-scale logarithmic derivative exactly
when the matching coefficient stops varying with the length of the middle
interval.  That residual variation is what the finite-size terms of
Eq.~\eqref{eq:joint_fit} are fitted to remove.  On the perturbative branch the
resulting derivative is the LL small-$x$ anomalous dimension.

The second-order ODE picture is heuristic, since the BFKL kernel is non-local; in the continuum, the power in Eq.~\eqref{eq:two_power_match} follows from the leading perturbative pole $\gamma(\aw)$ of $[1-\aw\chi(\gamma)]^{-1}$ in the Mellin representation, with residue $-1/(\aw\chi'(\gamma))$.

Equation~\eqref{eq:green_power} is a continuum matching statement, not an exact
identity for a finite principal section.  The finite claim tested below is more
modest: along sequences in which all three separations of
Eq.~\eqref{eq:three_scales} grow, the coefficient of $\log(N/N_0)$ in
$\log\GG$ approaches the perturbative solution of
Eq.~\eqref{eq:characteristic}.

Write
\begin{equation}
\gamma(\aw)=\sum_{r\ge1}\gamma_r\aw^r,
\qquad
C_r(D,N,N_0)\approx \gamma_r\log\frac{N}{N_0}.
\end{equation}
Choose
$D_N=\kappa N-1$ and $D_{2N}=2\kappa N-1$, so that the ratio of the upper
boundary to the observation cell is unchanged under $N\to2N$.  The paired
estimator is
\begin{equation}
\widehat\gamma_r(N,N_0\mid\kappa)
=
\frac{C_r(D_{2N},2N,N_0)-C_r(D_N,N,N_0)}{\log2}.
\label{eq:gamma_estimator}
\end{equation}
The name describes the construction.  The two calculations use observation
cells $N$ and $2N$, so the logarithm of the scale ratio increases by exactly
$\log2$.  At the same time, the final cell changes from $D_N=\kappa N-1$ to
$D_{2N}=2\kappa N-1$, which keeps the leading ultraviolet geometry $D/N$
fixed.  The doubling difference therefore measures the logarithmic slope
while comparing two systems with the same relative distance from the upper
boundary.

Taking the difference of two geometries removes the scale-independent
normalisations and isolates $\gamma_r$ up to finite-boundary corrections.

The perturbative branch is the one with $\gamma(0)=0$ and $\gamma'(0)=1$.  It
meets the reflected branch at
\begin{equation}
a_{\omega,c}=\frac{1}{4\log2},
\qquad
\gamma(a_{\omega,c})=\frac12.
\label{eq:branch_point}
\end{equation}
Because $\aw=\ab/\omega$, this is the same event as the upper end of the spectrum, the
largest continuum eigenvalue of Eq.~\eqref{eq:bfkl_edge_intro}.  The meeting
point is a branch point: it is the value of $\aw$ at which the two solutions
merge and the perturbative series for $\gamma(\aw)$ stops converging.  The
first singularity of the anomalous-dimension series and the upper end of the
fixed-coupling BFKL spectrum are thus two views of one event.

A direct test at finite $\aw$ uses
\begin{equation}
\gamma_{\rm eff}(\aw;N,N_0,\kappa)
=
\frac{
\log\GG_{2\kappa N-1,2N,N_0}(\aw)
-
\log\GG_{\kappa N-1,N,N_0}(\aw)}{\log2}.
\label{eq:gamma_eff}
\end{equation}
If Eq.~\eqref{eq:green_power} controls the long interval, this quantity must
approach the solution of Eq.~\eqref{eq:characteristic}.

\begin{figure}[ht]
	\centering
	\subfloat[The two continuum powers \label{fig:gammaeff_d}]{
		\includegraphics[width=0.48\textwidth]{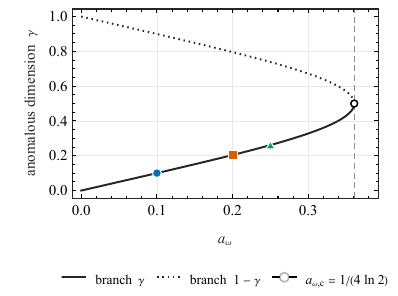}}
		\hfill
	\subfloat[Complete finite inverse matrix for $a_\omega = 0.1$\label{fig:gammaeff_a}]{%
		\includegraphics[width=0.48\textwidth]{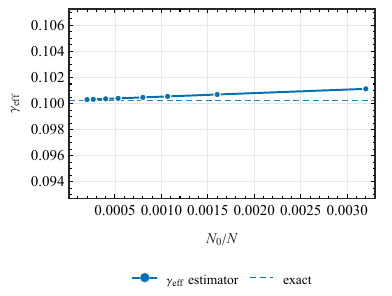}}
	\vspace{0.6em}
		\subfloat[ Complete finite inverse matrix for $a_\omega = 0.2$\label{fig:gammaeff_b}]{%
		\includegraphics[width=0.48\textwidth]{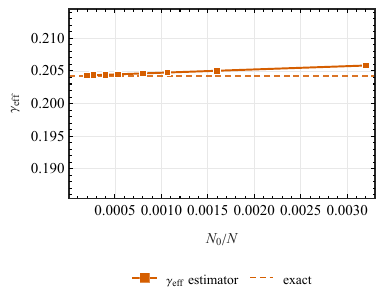}}
		\hfill
	\subfloat[ Complete finite inverse matrix for $a_\omega = 0.25$\label{fig:gammaeff_c}]{%
		\includegraphics[width=0.48\textwidth]{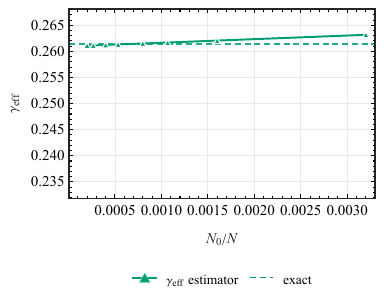}}\hfill

	\caption{Panel (a) solves Eq.~\eqref{eq:characteristic} and displays the two
		branches that meet at Eq.~\eqref{eq:branch_point}.  Panels (b)--(d) evaluate the
		complete inverse matrix through Eqs.~\eqref{eq:normalised_green} and
		\eqref{eq:gamma_eff}, with $N_0=20$, $\kappa=16$ and
		$\aw=0.10,0.20,0.25$; dashed lines are the continuum solutions of
		Eq.~\eqref{eq:characteristic}.  The largest scale-doubling comparison reaches
		$D=3\,199\,999$.  The complete inverse matrix approaches the same exponent selected by one interior row.}
	\label{fig:anomalous_checks}
\end{figure}

Fig.~\ref{fig:anomalous_checks} places the analytic branch structure and the complete finite-matrix test side by side.  Panel~(a) answers which of the two powers perturbation theory selects: the lower branch starts at $\gamma=0$ with
$\gamma\simeq\aw$ at small $\aw$, while the reflected branch starts near one, and they meet at $\gamma=1/2$.  Panels~(b)--(d) then ask whether the complete inverse matrix, with all powers of $\aw$ included, approaches that same lower branch when the scale hierarchy is enlarged.
In panels~(b)--(d) the horizontal coordinate is $N_0/N$, so the continuum limit lies
at the left edge.  Each coloured sequence approaches the dashed value obtained
from Eq.~\eqref{eq:characteristic}.  At the largest displayed sizes the
relative differences between $\gamma_{\rm eff}$ and the continuum branch are
about $0.059\%$, $0.051\%$ and $0.040\%$ for $\aw=0.10,0.20,0.25$.  Over the last
four sizes all three relative errors fall by roughly a factor of two whenever
$N$ is doubled, consistent with the $1/N$ family of corrections used in
Eq.~\eqref{eq:joint_fit}.  The approach is therefore uniform in $\aw$ over the
range shown; no loss of resolution between the two powers is seen at these
sizes, even closer to the branch point.  This test is stronger than a
comparison of a few perturbative coefficients, because it uses the complete
finite inverse matrix at non-zero $\aw$.  It remains a numerical test of the
matching argument, not a finite-dimensional proof.

\subsection{Perturbative coefficients}

Expanding Eq.~\eqref{eq:chi} around $\gamma=0$ gives
\begin{equation}
\chi(\gamma)
=
\frac1\gamma
+
2\sum_{k\ge1}\zeta(2k+1)\gamma^{2k},
\label{eq:chi_series}
\end{equation}
where $\zeta(s)=\sum_{n\ge1}n^{-s}$ is the Riemann zeta function; only its odd
integer values appear.  Rewriting the characteristic equation, Eq.~\eqref{eq:characteristic}, as
\begin{equation}
\gamma=\aw P(\gamma),
\qquad
P(z)=1+2\sum_{k\ge1}\zeta(2k+1)z^{2k+1},
\end{equation}
allows the inverse series to be obtained without solving the equation in closed
form.  By doing Lagrange inversion, we obtain
\begin{equation}
[\aw^r]\gamma(\aw)=\frac1r[z^{r-1}]P(z)^r,
\label{eq:lagrange}
\end{equation}
where $[z^n]$ means the coefficient of $z^n$ in a power series
\cite{FlajoletSedgewick:2009}.  The first terms are
\begin{equation}
\gamma(\aw)
=
\aw
+2\zeta_3\aw^4
+2\zeta_5\aw^6
+12\zeta_3^2\aw^7
+2\zeta_7\aw^8
+32\zeta_3\zeta_5\aw^9
+\cO(\aw^{10}).
\label{eq:gamma_series}
\end{equation}
The zero coefficients at orders two, three and five follow because $P(z)-1$
starts at $z^3$.  Single zeta values enter through the kernel, while products
such as $\zeta_3^2$ are created by the non-linear inversion.  This series is
the known LL small-$x$ result \cite{Jaroszewicz:1982,KotikovLipatov:2003}; here
it is used as a benchmark for the finite matrix.

\section{Numerical test of the three-scale limit}
\label{sec:numerics}

This section tests the continuum matching with the complete finite matrix.  The
key requirement is that different choices of source and ultraviolet geometry
must converge to one common anomalous-dimension coefficient. Throughout this
section a \emph{fit envelope} means the half-range of the results obtained when
the fit order and the fit window are varied. It measures sensitivity to the
finite-size extrapolation and is not a statistical standard deviation.

All calculations are matrix-free: the off-diagonal part of $H_D$ is Toeplitz, so matrix–vector products are computed with FFTs~\cite{CooleyTukey:1965} in $\mathcal O(D\log D)$; linear systems are solved by conjugate gradients~\cite{HestenesStiefel:1952} and eigenvalues by the Lanczos method~\cite{Lanczos:1950}.

The calculation uses ten observation scales
\begin{equation}
N=6250,12500,18750,25000,37500,50000,75000,100000,150000,200000
\label{eq:N_grid}
\end{equation}
and different geometry families depending on the perturbative order $r$. For $r = 1, \dots, 6$
\begin{equation}
(N_0,\kappa) = (10, 8), (20, 8), (10, 16), (20, 16), (10, 32), (20, 32), (10, 64), (20, 64),
\label{eq:geometry_families}
\end{equation}
and for $r = 7,8,9$
\begin{equation}
(N_0,\kappa) = (10, 512), (20, 512), (10, 1024), (20, 1024).
\label{eq:geometry_families2}
\end{equation}
With these families of geometries, the largest matrix used has $D \sim 4 \times 10^{8}$. The two sets are used for computational efficiency.
For each geometry $g$ in Eqs.~\eqref{eq:geometry_families} and
\eqref{eq:geometry_families2} the estimator is fitted separately to
\begin{equation}
	\widehat\gamma_r^{(g)}(N)
	=
	\gamma_r^{(g)}+
	\frac1N\sum_{j=0}^{k}c_{r,g,j}
	\left[\log\frac{N}{N_{0,g}}\right]^j,
	\qquad k=1,2,3,
	\label{eq:joint_fit}
\end{equation}
so each family has its own finite-size curve, because moving the source or the
upper boundary changes how the limit is approached.  The central test is that
the intercepts $\gamma_r^{(g)}$ of different geometries agree; our
estimate of $\gamma_r$ is their average over $g$, using the $k=3$ fit and all
ten values of $N$.  The fit envelope is obtained by repeating this average for
every $k\in\{1,2,3\}$ and for three fit windows, which keep all $N$ in
Eq.~\eqref{eq:N_grid} or drop the smallest one or two
($N\ge 6250$, $12500$, $18750$, always up to $N=200000$), and taking half the
range of these nine averages.

The horizontal coordinate in the next two figures is $10^5/N$, so larger
matrices lie farther to the left and the continuum limit is the intercept at
zero. The symbol with a vertical error bar at $10^5/N=0$ is the estimate of
$\gamma_r$ defined above (the $k=3$ intercept averaged over geometries) with
its fit envelope, while the dashed horizontal line is the known continuum
coefficient.

\begin{figure}[ht]
\centering
\includegraphics[width=0.80\textwidth]{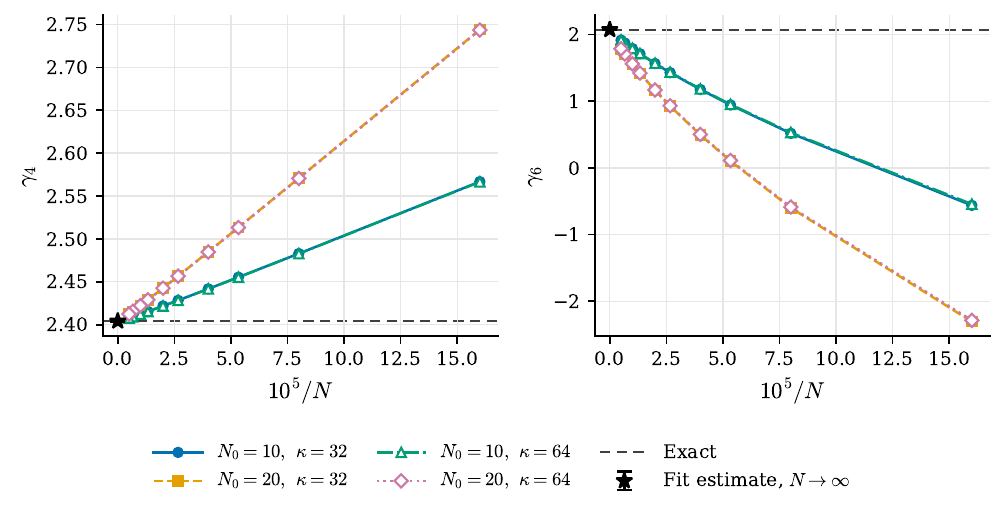}
\caption{Fourth- and sixth-order anomalous-dimension coefficients.  For
	clarity only the geometries with $\kappa=32,64$ of
	Eq.~\eqref{eq:geometry_families} are plotted; the extrapolated point uses all
	eight.  The differently placed finite lattices converge to a common value,
	close to $\gamma_4=2\zeta_3$ and $\gamma_6=2\zeta_5$.}
\label{fig:gamma4_extrapolation}
\end{figure}

Fig.~\ref{fig:gamma4_extrapolation} shows the fourth- and sixth-order coefficients.  All
four plotted geometries are already close to the analytic value, and the common
intercept is only weakly sensitive to the fit form.

\begin{figure}[ht]
\centering
\includegraphics[width=0.80\textwidth]{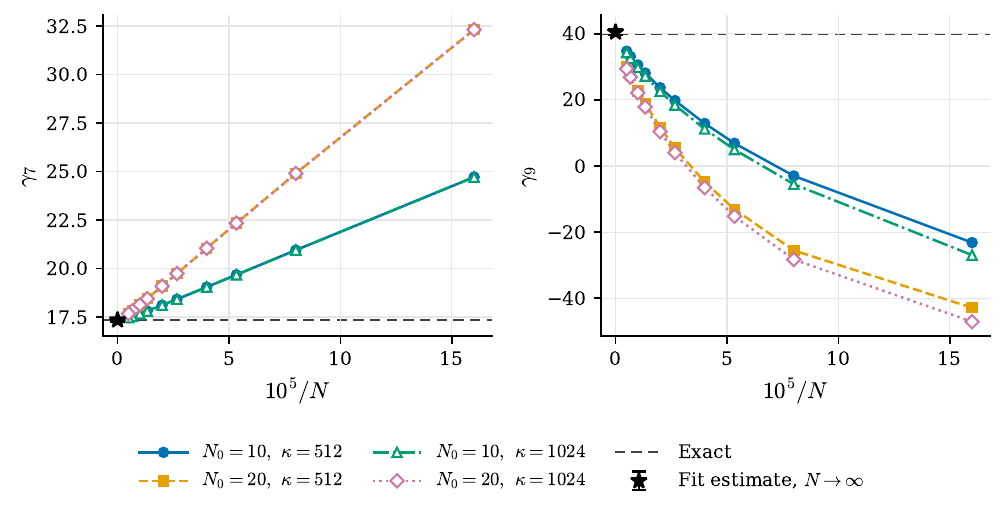}
\caption{Seventh and ninth-order coefficients from Eq.~\eqref{eq:joint_fit}. The large separation among the finite curves makes this the strongest test of the
three-scale extrapolation.  Their averaged intercept agrees with the known values
$\gamma_7=12\zeta_3^2$ and $\gamma_9=32\zeta_3\zeta_5$ within the displayed fit envelope.}
\label{fig:gamma7_extrapolation}
\end{figure}

Fig.~\ref{fig:gamma7_extrapolation} shows the seventh and the ninth coefficients.
The geometry families have large and visibly different finite-size drifts, yet the joint fit still brings them to values close to $12\zeta_3^2$ and $32\zeta_3\zeta_5$.  The vertical
range attached to the extrapolated point records the fit envelope.

\begin{table}[H]
	\centering
	\caption{Exact coefficients and direct matrix extrapolations.  The matrix
		column is the $k=3$ fit in Eq.~\eqref{eq:joint_fit}; the
		quoted uncertainty is the fit envelope defined at the start of this section. All numbers come from Eq.~\eqref{eq:joint_fit}
		with the geometries in Eqs.~\eqref{eq:N_grid},
		\eqref{eq:geometry_families} and \eqref{eq:geometry_families2}.}
	\label{tab:gamma_results}
	\begin{tabular}{@{}cccc@{}}
		\toprule
		$r$ & exact $\gamma_r$ & matrix extrapolation & status \\
		\midrule
		1 & $1$ & $1.00000058\pm0.00000084$ & resolved \\
		2 & $0$ & $(-7.2\pm6.6)\times10^{-7}$ & $1.1\times$ envelope from zero \\
		3 & $0$ & $(-2.8\pm7.4)\times10^{-6}$ & consistent with zero \\
		4 & $2\zeta_{3}=2.4041138$ & $2.40416\pm0.00069$ & resolved \\
		5 & $0$ & $-0.0003\pm0.0041$ & consistent with zero \\
		6 & $2\zeta_{5}=2.0738555$ & $2.0748\pm0.0011$ & resolved \\
		7 & $12\zeta_{3}^{2}=17.3392896$ & $17.340\pm0.087$ & resolved \\
		8 & $2\zeta_{7}=2.0166986$ & $1.98\pm0.37$ & resolved \\
		9 & $32\zeta_{3}\zeta_{5}=39.8862773$ & $40.46\pm0.78$ & resolved \\
		\bottomrule
	\end{tabular}
\end{table}

Two entries deserve comment.  At $r=2$ the extrapolation lies $1.1$ envelopes
from the exact zero, so the envelope slightly understates the spread there.  At
$r=8$ the exact coefficient is small compared with its neighbours, so a
comparable absolute envelope becomes an $18\%$ relative uncertainty, which makes
$\gamma_8$ the least precisely determined of the resolved coefficients.  The
highest-order resolved coefficient is
\begin{equation}
\gamma_9^{\rm matrix}=40.46\pm0.78_{\rm fit\ envelope},
\qquad
\gamma_9^{\rm LL}=39.8862773\ldots.
\end{equation}
The quoted uncertainty measures how stable the extrapolation is under changes
of fit order and fit window. It shrinks as the finite geometries
($N_0$, $N$, $D$) are enlarged, as Figs.~\ref{fig:gamma4_extrapolation} and
\ref{fig:gamma7_extrapolation} show.

\section{Fixed coupling: deformed spectra and Regge cut}
\label{sec:fixed_spectrum}

We now turn from scale evolution to the spectrum.  The goal is to see how a
finite set of poles turns into the fixed-coupling Regge cut, and to test how
that mechanism responds when real emission is deformed.

\subsection{A collinear Bessel deformation}

The collinear pole found in Section~\ref{sec:anomalous} (Eq.~\eqref{Eq: Collinear pole}) comes from emissions
ordered over a long range of transverse scales.  The LL hopping rule contains
no resummation of the double collinear logarithms that become important in that
region.  Here we multiply each real-emission transition by the known
Bessel resummation of their leading subset (two scale logarithms for each
extra power of the coupling) while keeping scale invariance intact
\cite{SabioVera:2005,ChachamisSabioVera:2016}.  The question asked is how much
of the spectral mechanism survives a physically motivated deformation. Note that the
result is not the full next-to-leading logarithmic (NLL) kernel
\cite{FadinLipatov:1998,CiafaloniCamici:1998,Salam:1998,CiafaloniColferaiSalam:1999}.

For two cells, define the logarithmic scale separation
\begin{equation}
\tau_{ij}=\log\frac{q_i^2}{q_j^2}=\log\frac{i}{j}.
\end{equation}
The real-emission multiplier is
\begin{equation}
b_{\ab}(\tau)
=
\frac{2J_1(\sqrt{2\ab}\,|\tau|)}{\sqrt{2\ab}\,|\tau|},
\qquad
b_{\ab}(0)=1,
\label{eq:bessel_factor}
\end{equation}
where $J_1$ is the Bessel function of the first kind.  Its expansion begins
\begin{equation}
\ab b_{\ab}(\tau)
=
\ab-\frac{\ab^2\tau^2}{4}+\frac{\ab^3\tau^4}{48}+\cdots,
\end{equation}
so nearby cells are unchanged at LL accuracy while widely separated jumps are
suppressed.  The factor starts at one, decreases, and eventually changes sign;
that sign change is why the positivity of the leading eigenvector, used later
for the LL running matrix, cannot be carried over to the Bessel model.  The
inertia argument used there needs only a positive coupling and is unaffected.
The finite physical matrix is
\begin{equation}
(H_D^B)_{ij}
=
(1-\delta_{ij})\frac{ b_{\ab}(\tau_{ij})}{|i-j|}
-2 h_{i-1}\delta_{ij}.
\label{eq:bessel_matrix}
\end{equation}
The virtual term keeps its LL value, and thus, the real-virtual cancellation still occurs, as the Bessel deformation approaches 1 when the cells are the same ($\tau_{ij} = 0$).  In Table~\ref{tab:bessel_edge} a comparison between the LL matrix and the Bessel deformed one is done.

Because the multiplier depends only on a ratio of scales, the continuum powers
remain eigenfunctions, and because it is even in $\tau=\log(i/j)$, the
reflection symmetry $\gamma\leftrightarrow1-\gamma$ is preserved.  The
eigenvalue (multiplied by the coupling) is
\begin{equation}
\omega_B(\gamma\mid\ab)
=
\ab\int_0^1\frac{\dd x}{1-x}
\left[b_{\ab}(\log x)(x^{\gamma-1}+x^{-\gamma})-2\right].
\label{eq:bessel_eigenvalue}
\end{equation}
On the symmetric line $\gamma=\tfrac12+i\nu$, setting $x=e^{-u}$ turns this
into the real integral
\begin{equation}
\omega_B\!\left(\frac12+i\nu\middle|\ab\right)
=
2\ab \int_0^\infty
\frac{b_{\ab}(-u)e^{-u/2}\cos(\nu u)-e^{-u}}{1-e^{-u}}\,\dd u,
\label{eq:bessel_real}
\end{equation}
whose numerator vanishes linearly at $u=0$, so the integral is finite.

\begin{figure}[ht]
	\centering
	\subfloat[Suppression of large jumps in $\tau = \log(q_i^2/q_j^2)$.]{\includegraphics[width = 0.47\textwidth]{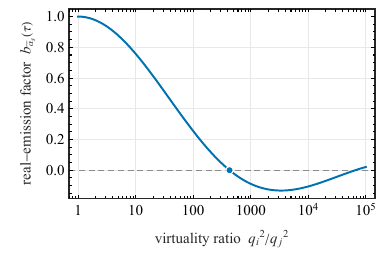}}
	\hspace{0.2cm}
	\subfloat[Change of the continuum spectral curve.]{\includegraphics[width = 0.43\textwidth]{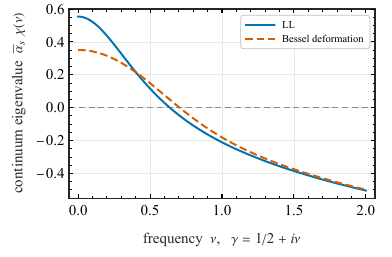}}
	\caption{Bessel deformation at $\bar\alpha_s=0.2$.  Panel (a) evaluates
		Eq.~\eqref{eq:bessel_factor}; the first zero occurs near the virtuality ratio
		$428$.  Panel (b) evaluates Eq.~\eqref{eq:bessel_real} and the undeformed LL
		curve.  Their values at $\nu=0$ are $0.3514$ and $0.5545$, respectively.  The
		deformation reshapes the spectrum rather than shifting it uniformly.}
	\label{fig:bessel_summary}
\end{figure}

Fig.~\ref{fig:bessel_summary} separates the local change of one transition
from its accumulated effect on the continuum spectrum.  Panel~(a) shows the
multiplier attached to a single real-emission jump: it equals one when the two
virtualities coincide, decreases as their ratio grows, and first vanishes near
a ratio of $428$ at $\ab=0.2$.  Panel~(b) integrates this modified jump weight
over all scale ratios for a continuum mode of frequency $\nu$; the value at
$\nu=0$ is the upper end of the spectrum.  The two curves cross at non-zero $\nu$, which
already shows that the deformation is not a uniform downward shift of the whole
spectrum.

The continuum edge is reduced from $0.554518$ to $0.351444$, a change of
$36.6\%$.  The first zero of the transition factor sits at logarithmic
separation $|\tau|\simeq6.06$, while a $D=4000$ integer lattice spans only
$\log D\simeq8.29$.  The deformation therefore acts over a substantial fraction
of the available logarithmic interval, which anticipates the slow finite-size
convergence seen in Table~\ref{tab:bessel_edge}.
\begin{table}[H]
\centering
\caption{Largest eigenvalues of $\ab H_D$ and $\ab H_D^B$.  Finite rows are
direct diagonalisations of Eqs.~\eqref{eq:main_matrix} and
\eqref{eq:bessel_matrix}; the continuum row evaluates
Eqs.~\eqref{eq:chi} and \eqref{eq:bessel_eigenvalue} at $\gamma=1/2$ and
$\ab=0.2$.}
\label{tab:bessel_edge}
\begin{tabular}{@{}rrrr@{}}
\toprule
$D$ & $\lambda_1^{\rm LL}$ & $\lambda_1^B$ & finite reduction \\
\midrule
100  & 0.331154 & 0.278779 & $15.8\%$ \\
500  & 0.378655 & 0.303249 & $19.9\%$ \\
2000 & 0.408415 & 0.315676 & $22.7\%$ \\
4000 & 0.420556 & 0.320132 & $23.9\%$ \\
continuum & 0.554518 & 0.351444 & $36.6\%$ \\
\bottomrule
\end{tabular}
\end{table}
The Rayleigh--Ritz principle tells us that for a real symmetric matrix $M$, the largest eigenvalue is the maximum of $x^{\mathsf T}Mx$ over all unit vectors $x$.  Both matrices in Table~\ref{tab:bessel_edge} are leading principal
submatrices of the corresponding matrices at size $D+1$, so every trial vector
at size $D$ is still available after one zero is appended.  Hence
$\lambda_{\max}(D)\leq\lambda_{\max}(D+1)$: the largest eigenvalue can only
grow with $D$.  This fixes the direction of the finite-size motion
but not its rate.

A useful identity explains why the eigenvectors can be less sensitive to the
Bessel deformation than the largest eigenvalue is.  For any vector $x$,
\begin{equation}
\langle x,H_Dx\rangle
=
\sum_{i=1}^{D}(h_{D-i}-h_{i-1})|x_i|^2
-
\frac12\sum_{i\ne j}\frac{|x_i-x_j|^2}{|i-j|}.
\label{eq:quadratic_form}
\end{equation}
The first term measures the imbalance between the two boundaries; the second is
never positive and penalises rapid variation across the lattice.  The leading
modes are therefore smooth, while their eigenvalues stay sensitive to the long
logarithmic interval and its boundaries.

When two nearly degenerate modes mix, comparing vectors one by one exaggerates the change.  Let $P_k^{\rm LL}$ and
$P_k^B$ be the orthogonal projectors onto the first $k$ modes of the two matrices and define
\begin{equation}
F_k=\frac1k\operatorname{Tr}(P_k^{\rm LL}P_k^B) =
\frac{1}{k}\sum_{m=1}^k\sum_{n=1}^k
\left|\langle\phi_m^{\rm LL}|\phi_n^B\rangle\right|^2,
\label{eq:subspace_overlap}
\end{equation}
where we wrote the projectors as $P_k^{\rm LL}=\sum_{m=1}^k|\phi_m^{\rm LL}\rangle\langle\phi_m^{\rm LL}|$
and similarly for $P_k^B$.

Thus $F_k$ averages all squared overlaps between the two leading sets of $k$
modes.  Equivalently, it is the mean fraction of an orthonormal basis of one
subspace that remains after projection onto the other.  It is unaffected by a
change of basis inside either subspace, satisfies $0\leq F_k\leq1$, and equals
one exactly when the two $k$-dimensional spaces coincide.  For $k=1$ it reduces
to the usual squared overlap of the two leading eigenvectors.
At $D=2000$ and $\bar\alpha_s=0.2$ we find $F_1=0.9885$ but $F_2=0.9989$.
The first number says that the individual leading vectors are close.  The even
larger second number says that almost all of the remaining change is a rotation
inside the leading two-mode plane rather than leakage into higher modes.

The position of a normalised eigenvector can be displayed by its cumulative
weight
\begin{equation}
W_n(m)=\sum_{i=1}^{m}|\phi_n(i)|^2.
\label{eq:cumulative_weight}
\end{equation}
For the continuum edge mode $\phi(i)\propto i^{-1/2}$, so multiplying a finite
eigenvector by $\sqrt{i}$ exposes any
oscillation or boundary distortion.

\begin{figure}[ht]
	\centering
	\subfloat[Compensated profiles of the first eigenfunctions.\label{fig:profiles}]{%
	\includegraphics[width=0.47\textwidth]{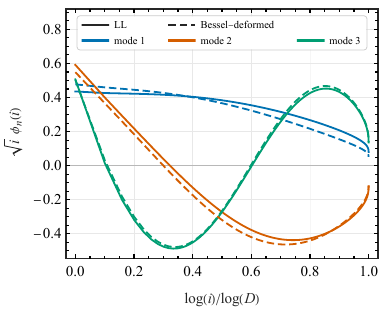}}\hfill
	\subfloat[Cumulative norm of the leading mode.\label{fig:cumulative}]{%
	\includegraphics[width=0.47\textwidth]{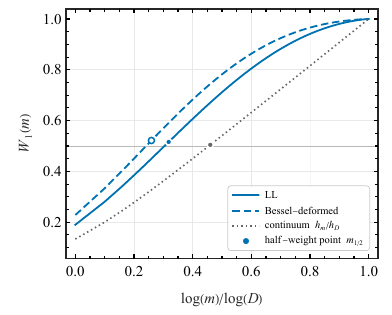}}
	\caption{Eigenvectors of the LL and Bessel-deformed matrices at $D=2000$ and
		$\bar\alpha_s=0.2$.  Panel~(a) shows the compensated profiles
		$\sqrt{i}\,\phi_n(i)$.  Panel~(b) plots the cumulative norm
		$W_1(m)=\sum_{i=1}^{m}|\phi_1(i)|^2$ against $\log m/\log D$.  Thus
		$W_1(m)$ is the fraction of the normalised leading mode contained in the
		first $m$ virtuality cells.  The dotted reference is
		$W_{\rm ref}(m)=h_m/h_D$, obtained from the continuum profile
		$\phi_{\rm ref}(i)\propto i^{-1/2}$.  The horizontal line $W_1=1/2$ defines
		the median cell: its first crossing moves from $m=12$ for the LL mode to
		$m=7$ after the Bessel deformation.  The leading two-mode subspace changes
		much less than the largest eigenvalue does.}
	\label{fig: norm evolution}
\end{figure}

Fig.~\ref{fig: norm evolution} makes the overlap numbers visible in cell
space.  Panel~(a) plots the first three compensated profiles: a perfectly
continuum-like edge profile would be horizontal after multiplication by
$\sqrt{i}$, so the remaining curvature and oscillation show finite-boundary
contributions, and the vertical offsets only keep the three modes apart.
Panel~(b) converts each normalised leading eigenvector into the cumulative
quantity $W_1(m)$ defined in Eq.~\eqref{eq:cumulative_weight}.  At a given
cell $m$, $W_1(m)$ is the fraction of the total squared norm contained in the
cells $1,\ldots,m$.  Since $q_m^2=m\Delta$, reaching a fixed cumulative
fraction at a smaller value of $m$ means that more of the eigenvector is
concentrated at lower virtuality.  We define the median cell $m_{1/2}$ as the
smallest cell for which
\[
W_1(m_{1/2})\geq\frac12.
\]
For the LL leading mode this occurs at $m_{1/2}=12$, whereas for the
Bessel-deformed leading mode it occurs at $m_{1/2}=7$.  The deformation
therefore shifts the norm of the leading mode towards the low-virtuality
cells.  These curves measure sensitivity to the deformation at fixed $D=2000$,
not convergence with matrix size.

\subsection{From poles to a cut}
At fixed coupling the eigenvalues are
$\lambda_p^{(D)}=\ab\varepsilon_p^{(D)}$ and the Green function is
\begin{equation}
G_D(\omega\mid i,j)
=
\sum_{p=1}^{D}\frac{\phi_p(i)\phi_p(j)}{\omega-\lambda_p^{(D)}}.
\label{eq:physical_pole_sum}
\end{equation}
If the weighted finite spectrum converges,
\begin{equation}
\sum_{p=1}^{D}\phi_p(i)\phi_p(j)
\delta(\lambda-\lambda_p^{(D)})
\longrightarrow
\rho_{ij}(\lambda),
\label{eq:weighted_measure_limit}
\end{equation}
where the Dirac delta represents a point mass at each finite eigenvalue, then
\begin{equation}
G(\omega\mid i,j)
=
\int\frac{\rho_{ij}(\lambda)}{\omega-\lambda}\,\dd\lambda.
\label{eq:stieltjes_green}
\end{equation}
This integral changes discontinuously when $\omega$ crosses the spectral
interval, and in the complex Mellin plane that discontinuity is the
fixed-coupling BFKL Regge cut.  If the levels crowd together while their source--observation
strengths vanish, their sum contributes nothing.  A cut needs both dense pole
positions and residues that survive.

Different densities address different questions. Frequency density: $\dd p/\dd\nu$ counts modes per unit logarithmic wave
frequency. Eigenvalue density: $|\dd p/\dd\lambda|$ counts modes per unit growth
rate. Weighted density:  $\rho_{ij}(\lambda)$ also contains the source and
observation overlaps, and only this  density enters the Green function.

The natural coordinate for the continuum modes is
\begin{equation}
t=\log\frac{q^2}{\mu^2},
\qquad
f_\nu(t)=e^{-t/2}e^{i\nu t}.
\end{equation}
Changing $\mu^2$ shifts $t$ by a constant but leaves the interval length
unchanged:
\[
t_{\max}-t_{\min}=\log\frac{q_{\max}^2}{q_{\min}^2}=\log D.
\]
A finite logarithmic interval cannot support arbitrary frequencies; the allowed
ones behave like standing waves on that interval.  Their positions can
be written as
\begin{equation}
\nu_p\log D+\vartheta_D(\nu_p)=\pi p,
\label{eq:phase_quantisation}
\end{equation}
where $\vartheta_D$ collects the extra phase produced by the lower and upper boundaries and
by the fact that the cells are uniform in $q^2$ rather than in $t$.
In Eq.~\eqref{eq:phase_quantisation} $\vartheta_D$ is extracted from the
spectrum.

We can map the largest eigenvalue $\varepsilon_1$ to a frequency $\nu_1$ using
\begin{equation}
\varepsilon_1(D)=\chi\!\left(\frac12+i\nu_1(D)\right).
\label{eq:edge_map}
\end{equation}
The data through $D=10^6$ give
\begin{equation}
\nu_1\left[\log D+c_{\rm Edge}+\cO((\log D)^{-1})\right]=\pi,
\qquad
c_{\rm Edge}=6.621\pm0.040.
\label{eq:edge_quantisation}
\end{equation}
Here the uncertainty measures the stability of the extrapolation under
reasonable fitting choices.  A \emph{fit form} is a chosen truncation of the
finite-size correction in powers of $1/\log D$, while a \emph{fit window} is
the subset obtained by changing the smallest matrix size.  If these
alternative fits give values between $c_{\min}$ and $c_{\max}$, we quote
\[
\Delta_{\rm fit}=\frac12(c_{\max}-c_{\min}).
\]
On the
symmetric line,
\begin{equation}
\begin{aligned}
\chi\!\left(\frac12+i\nu\right)
&=
4\log2
+2\sum_{k\ge1}(-1)^k\bigl(2^{2k+1}-1\bigr)
\zeta(2k+1)\nu^{2k}
\\
&=
4\log2-14\zeta_3\nu^2+62\zeta_5\nu^4+\cO(\nu^6),
\qquad |\nu|<\frac12,
\end{aligned}
\label{eq:edge_expansion}
\end{equation}
both coefficients following from expanding Eq.~\eqref{eq:chi} about
$\gamma=1/2$.  Keeping the leading quadratic term,
Eq.~\eqref{eq:edge_quantisation} implies
\begin{equation}
4\log2-\varepsilon_1(D)
\sim
\frac{14\zeta_3\pi^2}{(\log D+c_{\rm Edge})^2}.
\label{eq:edge_deficit}
\end{equation}

\begin{figure}[ht]
	\centering
	\subfloat[The first level obeys phase-shift quantisation.]{\includegraphics[width = 0.47\textwidth]{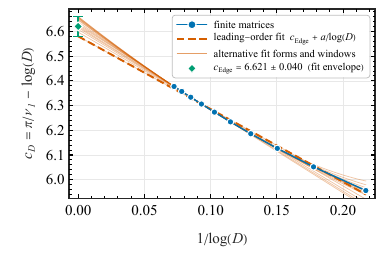}}
	\hspace{0.2cm}
	\subfloat[Edge deficit approaches a $1/(\log D)^2$ law.]{\includegraphics[width = 0.47\textwidth]{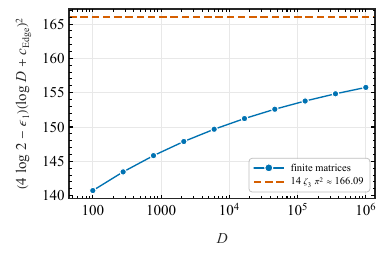}}
	\caption{Edge quantisation from Eqs.~\eqref{eq:edge_map}--\eqref{eq:edge_deficit}.
		Panel (a) extracts $c_{\rm Edge}$ from $c_D=\pi/\nu_1-\log D$; panel (b)
		removes the predicted $1/(\log D)^2$ dependence of the edge deficit.  The
		remaining offset is mostly the known quartic term in
		Eq.~\eqref{eq:edge_expansion}.  The dimensions are
		$D=100, 278, 774, 2154, 5995, 16681, 46416, 129155, 359381, 1000000$.  The same
		effective boundary length controls both panels.}
	\label{fig:edge_quantisation}
\end{figure}

Fig.~\ref{fig:edge_quantisation} tests this. In panel~(a) the combination $c_D=\pi/\nu_1-\log D$ would approach a constant if the first mode obeyed Eq.~\eqref{eq:edge_quantisation}; the data become nearly linear in $1/\log D$ and extrapolate to $c_{\rm Edge}=6.621\pm0.040$. Panel~(b) tests the consequence for the eigenvalue: multiplying the edge deficit by $(\log D+c_{\rm Edge})^2$ removes the leading size dependence. The scaled points stay below the dashed quadratic limit with a well understood offset. At $D=10^6$ one finds $\nu_1=0.1554$; retaining only the quadratic term of Eq.~\eqref{eq:edge_expansion} would give the scaled value $169.71$, while the quartic term $62\zeta_5\nu_1^4$ lowers the deficit by a further $9.2\%$, giving $154.06$; the sextic term raises this to $155.56$, close to the observed $155.77$. This  supports the $1/(\log D)^2$ law.

The individually resolved intermediate-frequency levels provide an independent test of the level density.  Here ``intermediate frequency'' means that we omit both the first few levels near $\nu=0$ and the far high-frequency tail, where neighbouring levels can no longer be separated numerically.  Defining
consecutive frequencies and a local effective length by
\begin{equation}
\varepsilon_p^{(D)}
=
\chi\!\left(\frac12+i\nu_p^{(D)}\right),
\qquad
L_{p,\mathrm{eff}}^{(D)}
=
\frac{\pi}{\nu_{p+1}^{(D)}-\nu_p^{(D)}},
\label{eq:bulk_map}
\end{equation}
the numerical curves follow
\begin{equation}
L_{\mathrm{eff}}^{(D)}(\nu)=\log D+\Phi(\nu)+\Delta_D(\nu),
\quad
\Delta_D(\nu)\longrightarrow0,
\quad
\Phi(\nu)\simeq\log\frac4\nu
\quad(1.5\lesssim\nu\lesssim7),
\label{eq:bulk_law}
\end{equation}
where $L_{\mathrm{eff}}^{(D)}(\nu)$ denotes the curve sampled by the
discrete values $L_{p,\mathrm{eff}}^{(D)}$.  The second relation is an
empirical law and it does not apply to the first
edge level.

\begin{figure}[ht]
	\centering
	\subfloat[Local effective length.]{\includegraphics[width = 0.47\textwidth]{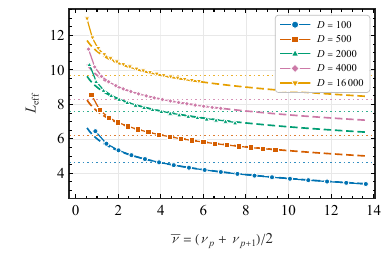}}
	\hspace{0.2cm}
	\subfloat[Collapse after subtracting $\log D$.]{\includegraphics[width = 0.47\textwidth]{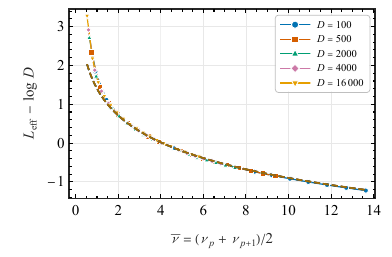}}
	\caption{Each finite eigenvalue is mapped to $\nu_p$ by
		Eq.~\eqref{eq:bulk_map}.  Panel (a) plots the local effective length
		$L_{p,\mathrm{eff}}$ for $D=100,500,2000,4000,16000$ and compares it with
		$\log D$ and the empirical form in Eq.~\eqref{eq:bulk_law}.  Panel (b) plots
		$L_{p,\mathrm{eff}}-\log D$; the collapse displays the common function $\Phi(\nu)$.
		Only the interval $1.5\lesssim\nu\lesssim7$, in which neighbouring levels are individually resolved, is fitted.  Subtracting $\log D$ isolates a common correction that depends on frequency but not on $D$. All the dashed lines are the respective fits.}
	\label{fig:mode_counting}
\end{figure}

Fig.~\ref{fig:mode_counting} shows why subtracting the logarithmic
volume is useful.  In panel~(a) the local effective length is not simply
$\log D$: it varies with the wave frequency, and the empirical curve
$\log(4D/\nu)$ follows this intermediate-frequency range much more closely.  Panel~(b)
subtracts $\log D$, after which data from five different matrix dimensions
collapse onto one common function $\Phi(\nu)$.

The points nearest $\nu=0$ do not follow this fit since their spacing is governed by the edge phase measured in
Fig.~\ref{fig:edge_quantisation}.  The collapse at larger $\nu$, on the other
hand, contains many levels and is the appropriate input for a
density-of-states limit, that is, a limit for the number of levels per unit
eigenvalue.  Keeping top-of-spectrum and intermediate-frequency regions apart avoids extracting an asymptotic law
from the wrong spectral region.

Dividing the unweighted counting density by the logarithmic volume gives
\begin{equation}
\frac1{\log D}\frac{\dd p}{\dd\nu}\longrightarrow\frac1\pi,
\end{equation}
and near the upper edge the quadratic term of Eq.~\eqref{eq:edge_expansion}
yields
\begin{equation}
\frac1{\log D}\left|\frac{\dd p}{\dd\varepsilon}\right|
\longrightarrow
\frac{1}{2\pi\sqrt{14\zeta_3(4\log2-\varepsilon)}}.
\label{eq:edge_density}
\end{equation}
The inverse square root is the density-of-states origin of the branch point.
With the physical eigenvalue $\lambda(\nu)=\ab\chi(1/2+i\nu)$, and provided the
external-mode weight is smooth and non-zero at $\nu=0$, the same expansion
gives
\begin{equation}
G(Y)\propto
\sqrt{\frac{\pi}{14\ab\zeta_3}}
\frac{e^{\omega_PY}}{\sqrt{Y}}.
\label{eq:largeY_cut}
\end{equation}
A single Regge pole would give a pure exponential.  The factor $Y^{-1/2}$ is due to the
continuum of nearby scale modes and is the signature of the
cut.

\section{Running coupling on the virtuality lattice}
\label{sec:running}

At fixed coupling, all virtuality cells interact with the same strength.  A
running coupling breaks that scale symmetry because the interaction becomes
weaker at large $q^2$ and must be specified separately in the low-virtuality
region.  The matrix Hamiltonian is useful because a coupling can be assigned to each cell,
the resulting real symmetric matrix is diagonalised and the role of the two
boundaries can be studied directly.

We study two complementary problems.  First, an infrared-finite coupling keeps
the low-scale cells in the matrix and remains regular through the scale at
which the one-loop expression would diverge.  A grid uniform in logarithmic
virtuality then makes it possible to count the states carried by the long
low-scale interval.  Second, following Lipatov~\cite{Lipatov:1986}, a one-loop construction removes the
low-scale region and replaces it by an imposed wall.  In that case the
decrease of the coupling creates a dynamical ultraviolet turning point above which a mode stops oscillating and begins to decay.  Table~\ref{tab:running_matrices} lists the matrices needed for these two questions.

\begin{table}[H]
\centering
\footnotesize
\caption{Matrices used in the running-coupling 	
	analysis.  The first two retain the regular low-virtuality region. The last replaces that region by a boundary so that the ultraviolet turning-point mechanism can be studied on its own.}
\begin{tabularx}{0.98\textwidth}{@{}>{\raggedright\arraybackslash}p{0.18\textwidth}>{\raggedright\arraybackslash}p{0.20\textwidth}>{\raggedright\arraybackslash}p{0.16\textwidth}X@{}}
\toprule
Matrix & Grid and coupling & Kernel & Purpose \\
\midrule
$\mathcal K^W_{D,\Delta}$ & uniform in $q^2$, infrared-finite & LL & refine the original Hamiltonian through the one-loop singular scale \\
$\mathsf T_{\delta,D}$ and $\mathcal K^W_{\delta,D}$ & uniform in $t$, fixed or infrared-finite & LL, with a frozen-Bessel comparison & reach long logarithmic intervals and count low-scale states \\
$\mathcal K_D^{\rm rc}$ & uniform in $q^2$, one-loop coupling with a lower wall & LL & isolate the turning-point and Airy mechanism \\
\bottomrule
\end{tabularx}
\label{tab:running_matrices}
\end{table}

Four coupling symbols are used below. These are the fixed value $\bar\alpha_s$, the
infrared-finite profile $\bar\alpha_W(q^2)$, its frozen limit
$\bar\alpha_\infty$ as $q^2\to0$, and the value $\bar\alpha_i$ in cell $i$.

\subsection{Implementing a finite infrared coupling in the matrix}
At one loop,
\begin{equation}
\bar\alpha_s(q^2)
=
\frac{1}{\bar\beta_0\log(q^2/\Lambda_{\rm QCD}^2)},
\qquad
\bar\beta_0=\frac{11N_c-2n_f}{12N_c},
\label{eq:running_one_loop}
\end{equation}
with $n_f$ the number of active quark flavours. The coupling weakens at large
$q^2$ (asymptotic freedom) but the same formula diverges at
$q^2=\Lambda_{\rm QCD}^2$.  That Landau singularity is a failure of the
one-loop formula, not a physical singularity of QCD.  To keep the low-virtuality
cells instead of discarding them, we use the infrared-finite form employed in
Ref.~\cite{HentschinskiSabioVeraSalas:2013}, based on Ref.~\cite{Webber:1998}.
The original motivation for this form came from studies of non-perturbative
power corrections to hard QCD observables.  In that framework, the
universality assumption is that several observables may depend on moments of
the same low-scale effective coupling, while the calculable coefficient that
multiplies each moment remains observable dependent
\cite{DokshitzerMarchesiniWebber:1996,Webber:1998}.  Thus ``universal''  
means that one common
infrared coupling can be tested across several observables.

Ref.~\cite{HentschinskiSabioVeraSalas:2013} used this parametrisation in a
small-$x$ deep-inelastic-scattering calculation in order to extend the
analysis into the low-$Q^2$ region.  There it was motivated by its consistency
with global studies of power corrections at low scales and by its rapid
return to ordinary perturbative running at high scales.  

\begin{figure}[ht]
\centering
\includegraphics[width=0.65\textwidth]{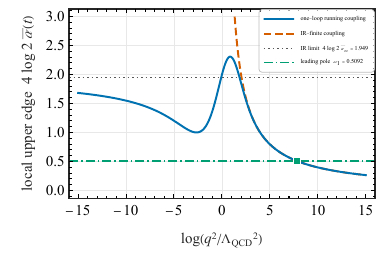}
\caption{Running-coupling profiles entering
Eqs.~\eqref{eq:running_one_loop} and \eqref{eq:webber_coupling_short}.  The
vertical quantity is the local LL upper edge $4\log2\,\bar\alpha(t)$.  The
one-loop curve has a Landau singularity at $t=0$; the Webber curve is finite,
freezes as $t\to-\infty$, and returns to one-loop running in the ultraviolet.
The horizontal line marks the leading pole of the one-loop-wall benchmark
discussed below.  The figure separates the running-coupling profile from the finite-matrix spectrum.}
\label{fig:running_profiles_short}
\end{figure}

With $\xi=q^2/\Lambda_{\rm QCD}^2$,
\begin{equation}
\bar\alpha_W(\xi)
=
\frac1{\bar\beta_0}
\left[
\frac1{\log\xi}
+
\frac{125(1+4\xi)}{(1-\xi)(4+\xi)^4}
\right].
\label{eq:webber_coupling_short}
\end{equation}
The two terms are separately singular at $\xi=1$, but their poles cancel, and
the resulting coupling satisfies
\begin{equation}
\bar\alpha_W(1)=\frac1{2\bar\beta_0},
\qquad
\bar\alpha_W(0)=\frac{125}{256\,\bar\beta_0},
\qquad
\bar\alpha_W(\xi)
=
\frac1{\bar\beta_0\log\xi}+\cO(\xi^{-4})
\quad(\xi\to\infty).
\label{eq:webber_limits_short}
\end{equation}
It therefore freezes to a finite constant at
small virtuality and returns quickly to one-loop running at large virtuality.
It is not monotonic: for $N_c=3$ and $n_f=4$, the values used in all running-coupling results below, it peaks near $\xi=2.38$. Fig.~\ref{fig:running_profiles_short}
shows the corresponding local profile $4\log2\,\bar\alpha(t)$.

On the original cells $q_i^2=i\Delta$, collect the couplings in a positive
diagonal matrix
\begin{equation}
A^W_{D,\Delta}
=
\diag\!\left[
\bar\alpha_W\!\left(\frac{\Delta}{\Lambda_{\rm QCD}^2}\right),
\ldots,
\bar\alpha_W\!\left(\frac{D\Delta}{\Lambda_{\rm QCD}^2}\right)
\right]
\label{eq:webber_A_short}
\end{equation}
and define the running Hamiltonian
\begin{equation}
\mathcal K^W_{D,\Delta}
=
(A^W_{D,\Delta})^{1/2}H_D(A^W_{D,\Delta})^{1/2}.
\label{eq:webber_matrix_short}
\end{equation}
Each hop carries half a coupling from the cell it leaves and half
from the cell it reaches, so its off-diagonal entry is the original weight
$1/|i-j|$ multiplied by the geometric mean of the two couplings, while the
diagonal is $-2\bar\alpha_W(i\Delta/\Lambda_{\rm QCD}^2)h_{i-1}$.  Splitting
the coupling this way keeps the matrix symmetric and preserves the local
real--virtual cancellation.  Because
$\bar\alpha_W$ is finite and positive for $q^2>0$, every entry stays
finite as cells cross $q^2=\Lambda_{\rm QCD}^2$ and move towards the origin.

The symmetric placement also has a simple local interpretation.  In a region
where the coupling varies slowly, a mode looks over a few neighbouring cells
like $e^{i\nu t}$.  Because the same factor
$\sqrt{\bar\alpha(t)}$ appears on both sides, the terms containing one
derivative cancel.  The leading local relation is therefore
\begin{equation}
\omega=\bar\alpha(t)\,\chi\!\left(\frac12+i\nu\right)
+\cO(\partial_t^2\bar\alpha).
\label{eq:symmetric_local_symbol_short}
\end{equation}
Thus the fixed-coupling characteristic function remains the local
starting point, while corrections begin only when the coupling changes
appreciably across the wavelength of the mode.

The same positive invertible matrix
multiplies $H_D$ on both sides. Such a rescaling can move the levels and
change their residues but cannot change how many are positive, negative or
zero.   Besides, the original Hamiltonian
remains visible inside the running scenario. If
$\mathcal K^W_{D,\Delta}\phi_n=\omega_n\phi_n$ and
$f_n=(A^W_{D,\Delta})^{1/2}\phi_n$, then
\begin{equation}
H_Df_n
=
\omega_n(A^W_{D,\Delta})^{-1}f_n,
\qquad
f_m^{\mathsf T}(A^W_{D,\Delta})^{-1}f_n=\delta_{mn}.
\label{eq:generalised_running_short}
\end{equation}
The running coupling does not replace the lattice Hamiltonian; it only changes the positive weight used to normalise its eigenvectors. The positive off-diagonal entries ensure that the leading eigenvector can be chosen strictly positive.  Computationally we apply
$(A^W)^{1/2}$, then the same fast-Fourier action of $H_D$ used at fixed
coupling, then $(A^W)^{1/2}$ again.

Because the coupling is regular everywhere, the continuum limit
\begin{equation}
\Delta\to0,
\qquad
N_0\Delta\to Q_0^2,
\qquad
N\Delta\to Q^2,
\qquad
D\Delta\to Q_{\max}^2
\label{eq:running_physical_refinement}
\end{equation}
can be taken first, and then the ultraviolet boundary $Q_{\max}^2$ sent to
infinity.  This differs from the one-loop-wall construction, studied in our last subsection, with fixed $t_0>0$,
where $\Delta=\Lambda_{\rm QCD}^2e^{t_0}$ is fixed and increasing $D$ enlarges
only the ultraviolet range.

\begin{figure}[ht]
\centering
\includegraphics[width=0.6\textwidth]{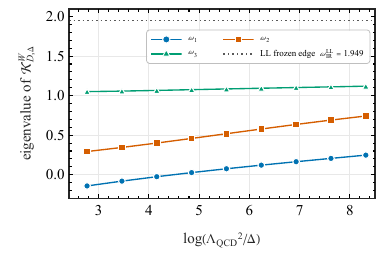}
\caption{Infrared-completed Hamiltonian Eq.~\eqref{eq:webber_matrix_short} on cells uniform in $q^2$.  The physical
upper scale is fixed at $Q_{\max}^2=256\Lambda_{\rm QCD}^2$ while
$\Delta/\Lambda_{\rm QCD}^2$ decreases from $2^{-4}$ to $2^{-12}$, so the
largest matrix has $D=1\,048\,576$.  The calculation uses a matrix-free
Lanczos method.  The horizontal reference is the LL edge of the frozen
infrared plateau.  The finite operator remains regular as the grid moves through the one-loop Landau scale.}
\label{fig:webber_q2_short}
\end{figure}

Fig.~\ref{fig:webber_q2_short} tests Eq.~\eqref{eq:webber_matrix_short}
directly on the original $q^2$ lattice, reducing the cell width from
$2^{-4}\Lambda_{\rm QCD}^2$ to $2^{-12}\Lambda_{\rm QCD}^2$ at fixed
$Q_{\max}^2=256\Lambda_{\rm QCD}^2$. The leading levels change smoothly as the grid is refined through the one-loop Landau scale. The higher levels move more, because refining the
grid also opens additional logarithmic length near the origin.

\subsection{Counting the states carried by the low-virtuality region}
The original discretised lattice is not numerically optimal when we operate with the variable  $t=\log(q^2/\Lambda_{\rm QCD}^2)$.  For the low-scale spectrum it is more convenient to introduce an alternative  matrix on a grid that is uniform in $t$,
\begin{equation}
	t_i=t_{\min}+i\delta,\qquad i=0,\dots,D-1,\qquad
	\delta=\frac{L_{\rm IR}}{D-1},
	\label{eq:log_grid_short}
\end{equation}
where $L_{\rm IR}=t_{\max}-t_{\min}$ is the length of the grid. Throughout this
subsection $t_{\max}$ is held fixed and $L_{\rm IR}$ is varied, with
$t_{\min}=t_{\max}-L_{\rm IR}$ adjusting accordingly. After removing $e^{-t/2}$, the fixed-coupling kernel takes
the symmetric form
\begin{equation}
	(\mathcal Lu)(t)
	=
	\int_0^\infty\!\dd s
	\left[
	\frac{u(t+s)+u(t-s)}{2\sinh(s/2)}
	-
	\frac{2u(t)}{e^s-1}
	\right],
	\label{eq:log_operator_short}
\end{equation}
where, as before, the two terms must be combined before $s\to0$, since their
$1/s$ singularities cancel. On a finite grid both terms are restricted to the
interval $[t_{\min},t_{\max}]$: real emissions reach only grid points, and the
virtual term becomes
\begin{equation}
	-u(t)\left[\int_0^{t-t_{\min}}\frac{\dd s}{e^s-1}
	+\int_0^{t_{\max}-t}\frac{\dd s}{e^s-1}\right].
	\label{eq:log_virtual_finite}
\end{equation}
Each integral in Eq.~\eqref{eq:log_virtual_finite} is logarithmically divergent
at $s=0$ on its own; only its combination with the real-emission term is finite,
and the discretisation below makes that cancellation explicit.
Discretising Eqs.~\eqref{eq:log_operator_short} and
\eqref{eq:log_virtual_finite} with the trapezoidal rule on the
grid~\eqref{eq:log_grid_short} gives a real symmetric matrix
$\mathsf T_{\delta,D}$ with off-diagonal entries
\begin{equation}
	(\mathsf T_{\delta,D})_{ij}
	=
	\frac{\delta}{2\sinh(\delta|i-j|/2)}
	\quad(i\ne j),
	\label{eq:log_offdiag_short}
\end{equation}
and diagonal entries
\begin{equation}
	(\mathsf T_{\delta,D})_{ii}
	=-\delta\left[S(i-1)+S(D-1-i)\right],
	\qquad
	S(N)=\sum_{r=1}^{N}\frac{1}{e^{r\delta}-1}.
	\label{eq:T_diag_exact}
\end{equation}
The off-diagonal entries depend only on $|i-j|$, but the diagonal varies near
the two ends of the grid, so $\mathsf T_{\delta,D}$ is Toeplitz only in the
bulk, in the same way as $H_D$ in Section~\ref{sec:finite}. Away from the boundaries, $S(N)\to S(\infty)$ up to exponentially small corrections, and an
Euler--Maclaurin expansion of $S(\infty)$ gives the bulk value
\begin{equation}
	(\mathsf T_\delta)_{ii}
	=
	2\psi(1)+2\log\delta-\frac{\delta}{2}+\frac{\delta^2}{72}+\cO(\delta^4).
	\label{eq:log_diagonal_asymptotic_short}
\end{equation}
The $2\log\delta$ term is the discrete counterpart of the short-distance
cancellation in Eq.~\eqref{eq:log_operator_short}: it compensates the
$\delta/\sinh(\delta r/2)\simeq 2/r$ growth of the off-diagonal entries near
the diagonal. The term $-\delta/2$ shifts the whole spectrum uniformly and
vanishes in the continuum limit.

 Dressing
the matrix by the coupling gives
\begin{equation}
\mathcal K^W_{\delta,D}
=
(A^W_{\delta,D})^{1/2}\mathsf T_{\delta,D}(A^W_{\delta,D})^{1/2},
\qquad
(A^W_{\delta,D})_{ii}=\bar\alpha_W(e^{t_i}).
\label{eq:webber_log_matrix_short}
\end{equation}

As $t\to-\infty$ the coupling approaches the positive constant
$\bar\alpha_\infty=\bar\alpha_W(0)$, so the deep low-virtuality region becomes
an approximately translation-invariant fixed-coupling medium.  A local mode
there has
\begin{equation}
\omega(\nu)=\bar\alpha_\infty
\chi\!\left(\frac12+i\nu\right),
\label{eq:frozen_dispersion_short}
\end{equation}
and since $\bar\alpha_\infty>0$, the sign of $\omega$ is the sign of $\chi$.
Positive states therefore occupy the frequency interval
\begin{equation}
0\le\nu<\nu_*,
\qquad
\chi\!\left(\frac12+i\nu_*\right)=0,
\qquad
\nu_*=0.6375.
\label{eq:nu_star_short}
\end{equation}
On a low-virtuality
interval of logarithmic length $L_{\rm IR}$ the two boundaries impose a
standing-wave condition
\begin{equation}
\nu_nL_{\rm IR}+\vartheta_{\rm end}(\nu_n)=\pi n,
\label{eq:ir_quantisation_short}
\end{equation}
whose boundary phase $\vartheta_{\rm end}$ shifts the count only by a bounded
amount.  The number of frequencies below $\nu_*$ is therefore
\begin{equation}
N_+(L_{\rm IR})
=
\frac{\nu_*}{\pi}L_{\rm IR}+\cO(1),
\qquad
\frac{\nu_*}{\pi}=0.2029.
\label{eq:universal_count_short}
\end{equation}
The coupling cannot change this slope at all.  The dressing in
Eq.~\eqref{eq:webber_log_matrix_short} multiplies $\mathsf T_{\delta,D}$ on both
sides by the same positive matrix, so by the inertia argument below
Eq.~\eqref{eq:webber_matrix_short} the number of positive states at a given
$L_{\rm IR}$ is that of $\mathsf T_{\delta,D}$ itself, for every positive
coupling profile.  Unlike the positivity of the leading eigenvector, this
argument requires only $A>0$ and therefore also covers the Bessel comparison
below.  The slope is thus a property of the kernel, while the constant absorbs the two boundaries and the finite transition region.

\begin{figure}[ht]
	\centering
	\subfloat[Individual levels move as the interval opens.]{\includegraphics[width = 0.47\textwidth]{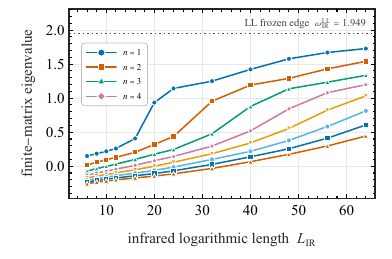}}
	\hspace{0.2cm}
	\subfloat[The residual stays bounded.]{\includegraphics[width = 0.47\textwidth]{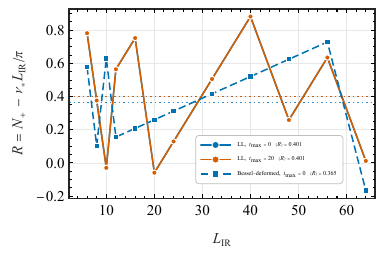}}
	\caption{Finite-matrix test of Eqs.~\eqref{eq:universal_count_short} and
		\eqref{eq:count_residual_short}. The plotted quantity is the integer state count minus the analytic linear term. Each kernel has its own slope removed, and the test is that each residual stays flat and bounded, since a
		linear drift would signal an incorrect predicted slope, and none is observed. All curves use $\delta=0.0125$. For $t_{\max}=20$ the grid extends into the perturbative running region $t>0$.}
	\label{fig:ir_count_short}
\end{figure}

Fig.~\ref{fig:ir_count_short} tests the stronger statement contained in
Eq.~\eqref{eq:universal_count_short}, namely that the residual left after
subtracting the analytic linear term,
\begin{equation}
R_K(L_{\rm IR};t_{\max})
=N_+^K(L_{\rm IR};t_{\max})
-\frac{\nu_*^K}{\pi}L_{\rm IR},
\qquad K=\hbox{LL},B,
\label{eq:count_residual_short}
\end{equation}
stays bounded.  For the LL kernel it stays within one level over a tenfold change in
$L_{\rm IR}$.  The two choices of $t_{\max}$ return the same integer count, as
the inertia argument requires; they differ in where the positive modes sit, not
in how many there are, so the second value tests the implementation rather than
the slope. The same test applied to the frozen Bessel deformation of Section~\ref{sec:fixed_spectrum}, evaluated at $\bar\alpha_s=\bar\alpha_\infty$, whose first positive zero is $\nu_*^B=0.7445$ and whose predicted slope is $\nu_*^B/\pi=0.2370$, shows that the linear law survives the deformation while the slope changes by about $17\%$, precisely because the positive-frequency interval of the kernel changes. A wrong analytic slope would make the residual drift linearly with $L_{\rm IR}$; the observed curves instead fluctuate around constant levels.

The height of the resulting band, the continuous range of eigenvalues
supported by the frozen low-scale region, is far less stable than its counting
law.  At $\bar\alpha_\infty=0.7031$ the LL edge is
$4\log2\,\bar\alpha_\infty=1.949$, whereas the Bessel-deformed edge is only
$0.5433$.  Such a large reduction is the reason why the LL value should not be
read as a reliable infrared intercept, while the state count is a much more
robust statement.

A coupling that freezes leaves an increasingly long, approximately translation-
invariant region in logarithmic virtuality, which is what makes it legitimate to
extend the grid to low $t$ at all.  The finite matrix then carries an increasing
number of positive states, with the slope set by the kernel.  This is different
from imposing a lower wall, which removes that region.  We now turn to that wall
problem as an alternative treatment of the origin of discrete running-coupling
poles.

\subsection{The one-loop wall as an alternative treatment of the infrared}

This subsection uses the last matrix in Table~\ref{tab:running_matrices}.  It
serves a different purpose from the infrared-completed matrices considered
above.  Here the region below a chosen scale is  replaced by
a lower wall.   This is the setting in which the
running-coupling BFKL cut was first related to a discrete set of positive
Pomeron poles, and in which the infrared boundary condition is
encoded by a phase
\cite{Lipatov:1986,KowalskiLipatovRoss:2014,RossSabioVera:2016}.

Choose cells
\begin{equation}
q_i^2=i\Lambda_{\rm QCD}^2e^{t_0},
\qquad
t_i=\log\frac{q_i^2}{\Lambda_{\rm QCD}^2}=t_0+\log i,
\qquad
\bar\alpha_i=\frac1{\bar\beta_0t_i},
\qquad t_0>0.
\label{eq:wall_cells_short}
\end{equation}
The first cell is at the logarithmic virtuality $t_0$, fixed to $t_0=4$
throughout this section; all
smaller virtualities have been replaced by a boundary condition.  We place the
coupling symmetrically around the fixed-coupling Hamiltonian,
\begin{equation}
\mathcal K_D^{\rm rc}=A_D^{1/2}H_DA_D^{1/2},
\qquad
A_D=\diag(\bar\alpha_1,\ldots,\bar\alpha_D).
\label{eq:wall_matrix_short}
\end{equation}
This is the same square-root ordering used in
Eq.~\eqref{eq:webber_matrix_short}.  It keeps the matrix real and symmetric,
so all finite eigenvalues are real.  The cell width is $\Delta=\Lambda_{\rm QCD}^2e^{t_0}$.  Increasing $D$ moves the upper boundary, $t_{\max}=t_0+\log D$.

Equation~\eqref{eq:characteristic} gave the fixed-coupling condition
$1=a_\omega\chi(\gamma)$.  In a slowly varying running background the same
argument is applied at each $t$, with
$a_\omega$ replaced by
\[
a_\omega(t)=\frac{\bar\alpha_s(t)}{\omega}.
\]
On the symmetric line $\gamma=\tfrac12+i\nu$ this gives
\begin{equation}
\omega
=
\bar\alpha_s(t)\,
\chi\!\left(\frac12+i\nu_\omega(t)\right),
\qquad\Longleftrightarrow\qquad
\chi\!\left(\frac12+i\nu_\omega(t)\right)
=
\bar\beta_0\omega t
\label{eq:wall_dispersion_short}
\end{equation}
for the one-loop coupling.  The quantity $\nu_\omega(t)$ is the wave
frequency in logarithmic virtuality.  This equation is the 
semiclassical form of the Hermitian running-coupling Green-function equation
used in Refs.~\cite{Lipatov:1986,KowalskiLipatovRoss:2014}.

For a given $t$, the largest value that the right-hand side of the 
spectral problem can support is obtained at $\nu=0$:
\begin{equation}
\omega_{\rm edge}(t)
=
\bar\alpha_s(t)\chi\!\left(\frac12\right)
=
\frac{4\log2}{\bar\beta_0t}.
\label{eq:wall_local_edge_short}
\end{equation}
This decreasing curve is the largest growth rate that the kernel can support
locally.  A positive eigenvalue
$\omega$ can be represented by a real frequency only while
$\omega<\omega_{\rm edge}(t)$.  The equality defines the turning point,
\begin{equation}
t_c(\omega)
=
\frac{4\log2}{\bar\beta_0\omega},
\label{eq:turning_point_short}
\end{equation}
where  $\nu_\omega(t_c)=0$.  Equivalently,
$a_\omega(t_c)=1/(4\log2)$, so the two solutions meet at
$\gamma=1/2$, exactly as they did at the fixed-coupling branch point in
Eq.~\eqref{eq:branch_point}.  Below $t_c$ the frequency is real and the mode
oscillates.  Above $t_c$ no real $\nu$ solves
Eq.~\eqref{eq:wall_dispersion_short}; the frequency becomes imaginary and the
physical solution decays.  The turning point is therefore a dynamical
ultraviolet boundary produced by asymptotic freedom.

\begin{figure}[ht]
\centering
\includegraphics[width=0.75\textwidth]{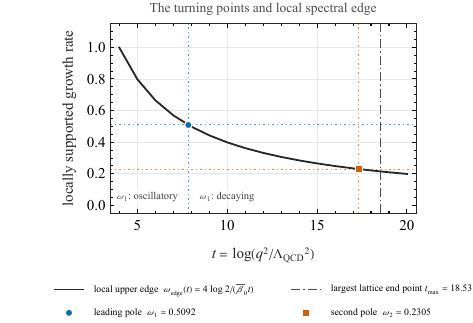}
\caption{Origin of the ultraviolet turning point.  The black curve is the
local upper edge in Eq.~\eqref{eq:wall_local_edge_short}.  The horizontal lines
are the two largest eigenvalues at $D=2.048\times10^6$; their intersections
with the edge give Eq.~\eqref{eq:turning_point_short}.  To the left of an
intersection a real local frequency exists and the mode oscillates.  To the
right it decays.  The final dotted line marks the position of the last lattice cell, $t_{\max}$. We have used $t_0 = 4$ to be far from the logarithmic divergence at $t = 0$.}
\label{fig:wall_turning_point_short}
\end{figure}

Fig.~\ref{fig:wall_turning_point_short} shows this construction for the two
largest eigenvalues at the largest matrix size used below.  The leading pole
meets the local edge at $t_c=7.84$, leaving a long decay interval before
$t_{\max}=18.53$.  The second pole turns only at $t_c=17.32$, so the available
decay interval is still short.  This indicates that the first pole
should be much less sensitive to the last cell than the second.

The  turning point fixes the
ultraviolet solution by requiring decay, while the wall at $t_0$ fixes
the infrared mixture of the two oscillatory solutions.  Their matching selects
a discrete set of values $\omega_n$.  This is the running-coupling analogue of
the fixed-volume phase condition in Eq.~\eqref{eq:phase_quantisation}. The
upper reflection is now generated by the turning point, whereas the infrared
phase remains a boundary input
\cite{KowalskiLipatovRoss:2014,RossSabioVera:2016}.

A useful measure of the oscillatory interval is the full-characteristic-function
action
\begin{equation}
s_\omega(t_0)
=
\int_{t_0}^{t_c(\omega)}\nu_\omega(t)\,\dd t,
\label{eq:wall_action_short}
\end{equation}
where $\nu_\omega(t)$ is obtained from
Eq.~\eqref{eq:wall_dispersion_short}, without replacing $\chi$ by its quadratic
approximation.  In the usual discrete-Pomeron description an infrared phase
completes this action to a quantisation condition.  The phase, and therefore
the pole positions, is a non-perturbative input \cite{RossSabioVera:2016}.

Close to $t_c$, both sides of the local relation can be expanded.  Using the
quadratic term of Eq.~\eqref{eq:edge_expansion},
$\chi(\tfrac12+i\nu)=4\log2-14\zeta_3\nu^2+\cdots$, and replacing the
frequency by the differential operator $\nu\to-i\partial_t$ gives
\begin{equation}
14\zeta_3u''(t)
+
\left(4\log2-\bar\beta_0\omega t\right)u(t)
\simeq0.
\label{eq:airy_short}
\end{equation}
With
\begin{equation}
\kappa_A
=
\left(\frac{\bar\beta_0\omega}{14\zeta_3}\right)^{1/3},
\qquad
z=\kappa_A(t-t_c),
\label{eq:airy_scale_short}
\end{equation}
this is $u_{zz}=zu$.  Its two solutions are the Airy functions
$\operatorname{Ai}(z)$ and $\operatorname{Bi}(z)$.  The first is selected
because it decays for $z>0$, whereas $\operatorname{Bi}(z)$ grows in the UV \cite{Lipatov:1986,KowalskiLipatovRoss:2014}.

The large-$z$ form
\begin{equation}
\operatorname{Ai}(z)
\sim
\frac{1}{2\sqrt\pi}\,z^{-1/4}
\exp\!\left(-\frac23z^{3/2}\right)
\qquad (z\to+\infty)
\label{eq:airy_decay_short}
\end{equation}
shows how the turning point removes sensitivity to a remote final cell.  If
$z_{\max}=\kappa_A(t_{\max}-t_c)$ is large, the amplitude reaching the final
cell is exponentially small; a reflected correction to the eigenvalue is
correspondingly
of order $\exp[-4z_{\max}^{3/2}/3]$.  At the largest displayed size,
$z_{\max,1}=2.95$ for the leading pole but only $z_{\max,2}=0.26$ for the
second.  Thus the leading mode has an ultraviolet decay region, while
the second is only beginning to acquire one.
\begin{figure}[ht]
	\centering
	\includegraphics[width=0.75\textwidth]{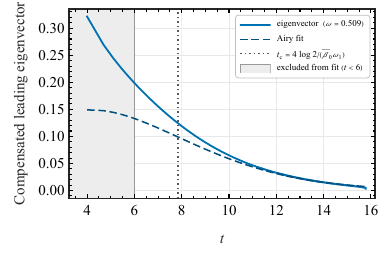}
	\caption{Airy profile of the leading one-loop-wall state at $D=131072$.
		The compensated eigenvector is compared with
		$\operatorname{Ai}[\kappa_A(t-t_c)]$ from
		Eqs.~\eqref{eq:airy_scale_short}--\eqref{eq:airy_decay_short}.  The shaded
		region $4\leq t\leq6$ is excluded from the fit because it is dominated by the
	 infrared wall.  The uniform-$t$ comparison has relative $L^2$ difference
		$0.237$ and correlation $0.995$.}
	\label{fig:wall_airy_profile_short}
\end{figure}
Fig.~\ref{fig:wall_airy_profile_short} compares the leading
eigenvector with this Airy solution.  The compensated eigenvector is the
lattice eigenvector multiplied by $\sqrt{i\bar\alpha_i}$.  Only one overall Airy amplitude is
fitted on $6\leq t\leq13$; the turning point is fixed by
Eq.~\eqref{eq:turning_point_short} and no horizontal shift is allowed.

\begin{figure}[ht]
\centering
\includegraphics[width=0.75\textwidth]{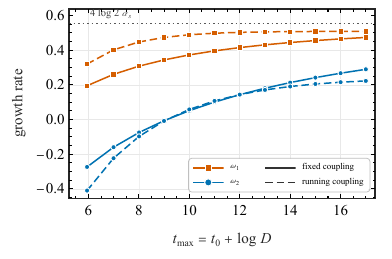}
\caption{Finite-size evolution of the two largest fixed- and running-coupling
levels.  The horizontal dotted line is the fixed-coupling continuum edge at
$\bar\alpha_s=0.2$.  The one-loop-wall leading pole reaches a plateau, whereas
the second pole is still rising at the largest $D$.  The contrast follows from
the long and short decay intervals displayed in
Fig.~\ref{fig:wall_turning_point_short}.}
\label{fig:wall_spectrum_short}
\end{figure}

Fig.~\ref{fig:wall_spectrum_short} compares the first two fixed-coupling
levels with the first two one-loop-wall eigenvalues.  At fixed coupling the
last cell is the ultraviolet wall, so moving it changes the standing-wave
length and both levels continue to drift towards the continuum edge.  With
running coupling the leading pole becomes nearly constant once a long Airy
tail lies between $t_c(\omega_1)$ and $t_{\max}$.  The second pole keeps moving
because its turning point still lies at or beyond the final cell at these
sizes.

This comparison also clarifies what the turning point does not do.  It removes
sensitivity to the distant ultraviolet cutoff, but it does not determine the
numerical pole position by itself.  The position still depends on how the
running coupling is defined  and on the boundary condition at
$t_0$.

At LL accuracy there is no unique prescription for the scale of the
coupling multiplying an off-diagonal transition.  We compare three symmetric
choices,
\begin{equation}
\bar\alpha_{ij}^{\rm geo}=\sqrt{\bar\alpha_i\bar\alpha_j},
\qquad
\bar\alpha_{ij}^{\rm hard}=\bar\alpha_{\max(i,j)},
\qquad
\bar\alpha_{ij}^{\rm soft}=\bar\alpha_{\min(i,j)},
\label{eq:wall_scale_choices_short}
\end{equation}
while keeping the same diagonal $-2\bar\alpha_i h_{i-1}$ in all three
matrices.  Since the coupling decreases with virtuality, the hard choice gives
the smallest real-emission weights and the soft choice the largest.  The
choices agree in the singular limit $i\to j$ and therefore preserve the local
real--virtual cancellation; they differ in finite, formally subleading terms.
The need to fix such scale and energy conventions consistently is one of the
central lessons of the next-to-leading BFKL and collinear-resummation analyses
\cite{FadinLipatov:1998,CiafaloniCamici:1998,Salam:1998,
CiafaloniColferaiSalam:1999}.

\begin{figure}[ht]
\centering
\includegraphics[width=0.75\textwidth]{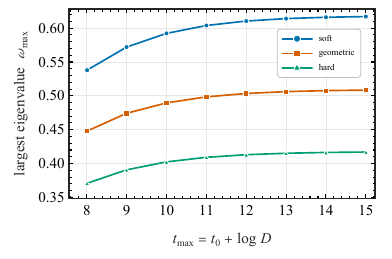}
\caption{Leading pole for the three running-scale choices in
Eq.~\eqref{eq:wall_scale_choices_short}.  Each sequence is already nearly flat
with $t_{\max}$, but the separation between the prescriptions is about $0.20$
at the largest common matrix size.  Ultraviolet stability does not imply
prescription independence.}
\label{fig:wall_scale_short}
\end{figure}

Fig.~\ref{fig:wall_scale_short} shows that this ambiguity is much larger than
the residual ultraviolet drift.  At $D=128000$ the three leading poles are
$\omega_1^{\rm hard}=0.417$, $\omega_1^{\rm geo}=0.509$,
$\omega_1^{\rm soft}=0.618$.  The spread, rather than the final digit of any
one curve, is the physical message of this LL model.

Some conclusions follow. First, the one-loop decrease of the coupling creates a
dynamical ultraviolet turning point and an Airy decay tail.  Second, once that
tail fits inside the finite lattice, the leading pole is genuinely insensitive
to the final cell.  Third, its numerical position is still strongly dependent on
the running-scale and infrared-boundary prescriptions.  The first two statements
concern the turning-point mechanism; the third prevents the stable value $0.509$
from being interpreted as a scheme-independent QCD prediction.

The two running-coupling calculations answer different questions.
The infrared-completed matrix shows what happens when the regular low-scale cells are kept. As expected, we found that the residual of the positive-state count oscillates around a mean value as the logarithmic length grows.  The one-loop-wall matrix instead isolates the traditional turning-point mechanism: ultraviolet decay makes the leading pole insensitive
to the final cell, while the running-scale and infrared-wall choices still set its numerical position.  In both calculations the finite matrix is the exact
starting point; continuum and Airy arguments are used only to interpret the
observed spectrum.

\section{Conclusions}
\label{sec:conclusions}

Replacing the continuous virtuality variable by $D$ cells turns leading-logarithmic
BFKL evolution into a single real symmetric matrix, Eq.~\eqref{eq:main_matrix}.
Its powers, inverse and spectral decomposition reduce the evolution to
finite-dimensional linear algebra.
A running coupling prescription can be inserted by the symmetric dressing of
Eq.~\eqref{eq:webber_matrix_short} in the same mathematical framework.

Acting with the matrix on the space of functions of the form $i^{\gamma-1}$ reproduces $\chi(\gamma)$ of
Eq.~\eqref{eq:chi} up to corrections shown in Eq.~\eqref{eq:row_error}. They correspond to the unresolved interval below the first cell, the
omitted emissions above the last cell  and those due to having sums instead of integrals.  Because they scale in different ways, the upper-tail correction can be separated
from the other two, which is what Figs.~\ref{fig:row_upper_tail} and
\ref{fig:row_lower_spacing} do.

Matching the powers related by $\gamma\leftrightarrow1-\gamma$ gives what we call the
source-to-observation behaviour $(N/N_0)^\gamma$ of
Eq.~\eqref{eq:green_power}.  We numerically test this form with the paired estimator of
Eq.~\eqref{eq:gamma_estimator}, fitted to the form in Eq.~\eqref{eq:joint_fit}.
In this way we are able to reproduce the coefficients of
Eq.~\eqref{eq:gamma_series} up to order nine, see Table~\ref{tab:gamma_results}.  The procedure
extends in principle to higher orders, but the extrapolation
uncertainty and the required matrix size grow quickly.

At fixed coupling, in logarithmic
virtuality space, one can interpret the finite levels in the context of standing waves.
The first frequency is governed by an effective length $\sim \log D+ 6.62$, Eq.~\eqref{eq:edge_quantisation}.
This implies that the distance to the asymptotic largest eigenvalue falls as $1/(\log D)^2$ (Eq.~\eqref{eq:edge_deficit}).
Away from the top of the spectrum, in the intermediate-frequency range, the
level density is $\log D/\pi$ per unit $\nu$. Near the top of the spectrum, at
small $\nu$, Eq.~\eqref{eq:edge_density} gives the inverse square root that is
the density-of-states origin of the branch point. It produces the $Y^{-1/2}$
factor of Eq.~\eqref{eq:largeY_cut} provided the external-mode weight is smooth
and non-zero at $\nu=0$.

The Bessel factor of Eq.~\eqref{eq:bessel_factor}, introduced to improve the kernel in collinear regions,
suppresses large changes in
scale while preserving scale invariance and the
$\gamma\leftrightarrow1-\gamma$ symmetry.  It lowers the continuum largest eigenvalue (Table~\ref{tab:bessel_edge}).
Interestingly, it barely changes the space spanned by the leading eigenvectors
(Eq.~\eqref{eq:subspace_overlap}).

We compared two treatments of low virtualities with a running coupling. In the
first, the low-scale cells are kept and the coupling freezes there with the infrared-finite profile of Eq.~\eqref{eq:webber_coupling_short}, motivated by studies of power corrections
to hard QCD observables. The
matrix is then well-defined at $q^2=\Lambda_{\rm QCD}^2$.  A grid of
logarithmic length $L_{\rm IR}$ supports standing waves spaced by
$\pi/L_{\rm IR}$ in frequency, and a wave grows with rapidity only if its
frequency lies below the first zero of $\chi$, at $\nu_*=0.6375$. The number of
growing (positive) states therefore rises linearly with $L_{\rm IR}$,
Eq.~\eqref{eq:universal_count_short}, with slope $\nu_*/\pi=0.2029$: about one
new state each time the virtuality range grows by a factor of $138$. In the
leading-logarithmic kernel the coupling cannot alter this count, because a
positive coupling changes how fast each state grows but not whether it grows;
the boundaries shift the count only by a bounded amount. How fast the states
grow is far less robust than how many there are. Under the Bessel deformation
the top of the low-scale band falls from $4\log 2\,\bar\alpha_\infty=1.949$ to
$0.5433$, while the slope changes by only about $17\%$.

In the second treatment a wall replaces the low-scale cells and the one-loop
coupling keeps decreasing with virtuality. The largest growth rate allowed
locally also decreases, Eq.~\eqref{eq:wall_local_edge_short}. A mode of
growth rate $\omega$ can oscillate only up to a turning point,
Eq.~\eqref{eq:turning_point_short}, above which it decays as an Airy function,
Eq.~\eqref{eq:airy_decay_short}. Once this decay fits inside the lattice, almost
no amplitude reaches the last cell, and moving it no longer changes the
pole. At the largest lattice used with the wall, $D=2.048\times10^6$, the Airy distance between the
turning point and the last cell is $z_{\max}=2.95$ for the leading pole but only
$0.26$ for the second. This is the contrast seen in
Fig.~\ref{fig:wall_spectrum_short}: the first pole flattens while the second
keeps rising.
The value of the leading pole is still not a prediction: it depends on the
running-scale prescription and on the infrared phase at the wall.

All the kernels used here are LL, except for the Bessel factor,
which resums the leading double logarithms of the scale ratio but is not the
full next-to-leading-logarithmic (NLL) kernel
\cite{FadinLipatov:1998,CiafaloniCamici:1998,Salam:1998,CiafaloniColferaiSalam:1999}.
Natural extensions of our work are to include further NLL terms and to adapt the
discretisation to the non-forward case relevant for diffraction. The formalism
can also be used to study quantum-information quantities when the running
coupling or higher-order (mostly collinear) corrections are included.

\section*{Acknowledgments}
We thank G. Chachamis and D. Vaccaro for discussions.  We acknowledge support from the Spanish Agencia Estatal de
Investigaci\'on (grants PID2022-142545NB-C22; IFT Centro de
Excelencia Severo Ochoa CEX2020-001007-S, funded by
MCIN/AEI/10.13039/501100011033 and by ERDF A way of making Europe).  We 
acknowledge the European Union Horizon 2020 programme (No.~824093).
{

}

\end{document}